\documentclass[useAMS,usenatbib]{mn2e}

\usepackage{amsmath}
\usepackage{graphicx}
\usepackage{hyperref}
\usepackage{upgreek}

\newcommand{\kms}{km s$^{-1}$}
\newcommand{\tex}{$T_{\text{ex}}$}
\def\jnl@style{\it}
\def\aaref@jnl#1{{\jnl@style#1}}
\def\aaref@jnl#1{{\jnl@style#1}}

\def\aj{\aaref@jnl{AJ}}                   % Astronomical Journal
\def\araa{\aaref@jnl{ARA\&A}}             % Annual Review of Astron and Astrophys
\def\apj{\aaref@jnl{ApJ}}                 % Astrophysical Journal
\def\apjl{\aaref@jnl{ApJ}}                % Astrophysical Journal, Letters
\def\apjs{\aaref@jnl{ApJS}}               % Astrophysical Journal, Supplement
\def\ao{\aaref@jnl{Appl.~Opt.}}           % Applied Optics
\def\apss{\aaref@jnl{Ap\&SS}}             % Astrophysics and Space Science
\def\aap{\aaref@jnl{A\&A}}                % Astronomy and Astrophysics
\def\aapr{\aaref@jnl{A\&A~Rev.}}          % Astronomy and Astrophysics Reviews
\def\aaps{\aaref@jnl{A\&AS}}              % Astronomy and Astrophysics, Supplement
\def\azh{\aaref@jnl{AZh}}                 % Astronomicheskii Zhurnal
\def\baas{\aaref@jnl{BAAS}}               % Bulletin of the AAS
\def\jrasc{\aaref@jnl{JRASC}}             % Journal of the RAS of Canada
\def\memras{\aaref@jnl{MmRAS}}            % Memoirs of the RAS
\def\mnras{\aaref@jnl{MNRAS}}             % Monthly Notices of the RAS
\def\pra{\aaref@jnl{Phys.~Rev.~A}}        % Physical Review A: General Physics
\def\prb{\aaref@jnl{Phys.~Rev.~B}}        % Physical Review B: Solid State
\def\prc{\aaref@jnl{Phys.~Rev.~C}}        % Physical Review C
\def\prd{\aaref@jnl{Phys.~Rev.~D}}        % Physical Review D
\def\pre{\aaref@jnl{Phys.~Rev.~E}}        % Physical Review E
\def\prl{\aaref@jnl{Phys.~Rev.~Lett.}}    % Physical Review Letters
\def\pasp{\aaref@jnl{PASP}}               % Publications of the ASP
\def\pasj{\aaref@jnl{PASJ}}               % Publications of the ASJ
\def\qjras{\aaref@jnl{QJRAS}}             % Quarterly Journal of the RAS
\def\skytel{\aaref@jnl{S\&T}}             % Sky and Telescope
\def\solphys{\aaref@jnl{Sol.~Phys.}}      % Solar Physics
\def\sovast{\aaref@jnl{Soviet~Ast.}}      % Soviet Astronomy
\def\ssr{\aaref@jnl{Space~Sci.~Rev.}}     % Space Science Reviews
\def\zap{\aaref@jnl{ZAp}}                 % Zeitschrift fuer Astrophysik
\def\nat{\aaref@jnl{Nature}}              % Nature
\def\iaucirc{\aaref@jnl{IAU~Circ.}}       % IAU Cirulars
\def\aplett{\aaref@jnl{Astrophys.~Lett.}} % Astrophysics Letters
\def\apspr{\aaref@jnl{Astrophys.~Space~Phys.~Res.}}
\def\bain{\aaref@jnl{Bull.~Astron.~Inst.~Netherlands}} 
\def\fcp{\aaref@jnl{Fund.~Cosmic~Phys.}}  % Fundamental Cosmic Physics
\def\gca{\aaref@jnl{Geochim.~Cosmochim.~Acta}}   % Geochimica Cosmochimica Acta
\def\grl{\aaref@jnl{Geophys.~Res.~Lett.}} % Geophysics Research Letters
\def\jcp{\aaref@jnl{J.~Chem.~Phys.}}      % Journal of Chemical Physics
\def\jgr{\aaref@jnl{J.~Geophys.~Res.}}    % Journal of Geophysics Research
\def\jqsrt{\aaref@jnl{J.~Quant.~Spec.~Radiat.~Transf.}}
\def\memsai{\aaref@jnl{Mem.~Soc.~Astron.~Italiana}}
\def\nphysa{\aaref@jnl{Nucl.~Phys.~A}}   % Nuclear Physics A
\def\physrep{\aaref@jnl{Phys.~Rep.}}   % Physics Reports
\def\physscr{\aaref@jnl{Phys.~Scr}}   % Physica Scripta
\def\planss{\aaref@jnl{Planet.~Space~Sci.}}   % Planetary Space Science
\def\procspie{\aaref@jnl{Proc.~SPIE}}   % Proceedings of the SPIE

\title{Deuteration in infrared dark clouds}

\author[M. Lackington et al.]{Matias Lackington$^{1}$, Gary A. Fuller$^{1,2}$,  Jaime E. Pineda$^{3}$, Guido Garay$^{4}$,
\newauthor Nicolas Peretto$^{5}$, and Alessio Traficante$^{1}$  \\
$^{1}$Jodrell Bank Centre for Astrophysics, Alan Turing Building, School of Physics and Astronomy, The University of Manchester,\\ Oxford Road, Manchester, M13 9PL, UK\\
$^{2}$UK ALMA Regional Center Node, Jodrell Bank Centre for Astrophysics, School of Physics and Astronomy,\\ The University of Manchester, Manchester, M13 9PL, UK\\
$^{3}$Max-Planck-Institut f\"{u}r extraterrestrische Physik, 85748 Garching, Germany\\
$^{4}$Departamento de Astronom\'{i}a, Universidad de Chile, Casilla 36-D Santiago, Chile\\
$^{5}$School of Physics and Astronomy, Cardiff University, Queen's buildings, Cardiff CF24 3AA, UK
}

\begin{document}

\date{Accepted 2015 October 8. Received 2015 October 7; in original form 2014 July 24}

\pubyear{2015}
\pagerange{001-015}

\maketitle

\begin{abstract}

Much of the dense gas in molecular clouds has a filamentary structure but the detailed structure and evolution of this gas is poorly known. We have observed 54 cores in infrared dark clouds (IRDCs) using N$_2$H$^+$ (1--0) and (3--2) to determine the kinematics of the densest material, where stars will form. We also observed N$_2$D$^+$ (3--2) towards 29 of the brightest peaks to analyse the level of deuteration which is an excellent probe of the quiescent of the early stages of star formation. There were 13 detections of N$_2$D$^+$ (3--2). This is one of the largest samples of IRDCs yet observed in these species. The deuteration ratio in these sources ranges between 0.003 and 0.14. For most of the sources the material traced by N$_2$D$^+$ and N$_2$H$^+$ (3--2) still has significant turbulent motions, however three objects show subthermal N$_2$D$^+$ velocity dispersion. Surprisingly the presence or absence of an embedded 70$\upmu$m source shows no correlation with the detection of N$_2$D$^+$ (3--2), nor does it correlate with any change in velocity dispersion or excitation temperature. Comparison with recent models of deuteration suggest evolutionary time-scales of these regions of several freefall times or less.
 
\end{abstract}


\section{Introduction}

Through their radiative, kinetic and chemical feedback into the interstellar medium, massive stars play a dominant role in the shaping and evolution of galaxies. However, the formation process and early evolution of high-mass stars is not well understood. Several factors contribute to this: (1) high-mass stars are intrinsically much rarer than their low-mass counterparts implying that they are generally observed at larger distances; (2) high-mass stars are rarely (if ever) formed in isolation, but in the complex environment of clusters and (3) high-mass stars, and in particular their youngest phases, are short lived. In addition, once formed, the feedback from young massive stars rapidly reshapes the regions in which they form.  Young massive stars are therefore typically seen in a complex environment involving gravitational interaction, winds, outflows and ionizing radiation. Thus the study of high-mass star from both observational and theoretical point of view is much more complex than that of low-mass star formation, with little consensus on the key mechanisms \citep{Zinnecker2007}.  To make progress it is crucial to identify massive star formation regions in their earliest stages, when the initial conditions and first stages of evolution can be studied.  Capturing these rare phases requires a large initial sample of massive, quiescent regions which could, or are just starting to, form massive stars.

Infrared dark clouds (IRDCs) are defined as regions with significant extinctions viewed against the bright, diffuse mid-infrared galactic background. Such regions were first detected by the \textit{Infrared Space Observatory} (\textit{ISO}) and then by the \textit{Midcourse Space Experiment} \citep[\textit{MSX};][]{ Perault1996, Egan1998, Hennebelle2001}, and are ubiquitous throughout the Galaxy \citep{Simon2006a}. Previous studies show that their molecular material has low temperatures ($<$25 K), high column densities ($\sim 10^{23}--10^{25}$ cm$^{-2}$), and high-volume densities ($>$10$^5$ cm$^{-3}$; \citealt{Egan1998}, \citealt{Carey1998,Carey2000}, \citealt{Peretto2009,Peretto2010a}). IRDCs have been described as possible precursors to the initial stages of massive star formation \citep{Rathborne2006}. Studies have shown \citep[e.g.][]{Teyssier2002,Pillai2006,Rathborne2006} that IRDCs are cold, and contain dense structures with sizes and masses comparable to warmer cluster-forming clumps and as well as compact cores with sizes, masses, and densities comparable to high-mass star-forming hot cores. Thus, IRDCs are a great starting place to search for the most quiescent and earlier stages of massive star formation.

Deuterium-containing molecular species allow us to study the quiescent and cold material in a molecular cloud. The deuterated species preferentially trace the coldest parts of dense cores (i.e. single star progenitors), and have been shown to be an important tool to determine the degree of evolution in low-mass cores \citep{Caselli2002,Crapsi2005,Emprechtinger2009} and high-mass cores \citep{Fontani2011,Pillai2012}. Chemical models \citep[e.g.][]{Kong2015} show increased deuterium enhancement with time in these cold and dense regions. In the prestellar core phase, the deuteration fractionation increases and CO freezes out on to grains as the core contracts, reaching maximum when star formation begins. In the prestellar phase the C-bearing species are therefore depleted towards the densest regions, while N-bearing species are good tracers of the total column density even at high volume densities. After the formation of a protostar, the deuteration fraction decreases as the protostar evolves, heats its environment and the CO is released from the dust grains. So a high deuteration fraction in a protostellar core indicates the youth of the protostar, while a high deuteration fraction in a pre-stellar core indicates that the core may soon collapse into a protostar. Thus, the deuteration ratio is an excellent probe of the early stages of star formation \cite[e.g.][]{Pineda2013}. A systematic study of deuteration towards high-mass star-forming regions will likely unveil a range of objects in different evolutionary stages, and will pinpoint the most likely candidates for the most quiescent regions.

In this paper, we use N$_2$H$^+$ and its deuterated counterpart N$_2$D$^+$ to study over 50 targets which are thought to be regions of massive star formation. In Section 2 of this paper, we present the source selection, data, and data reduction used in this study. Section 3 presents the analysis of the reduced data, and we show the derived parameters such as the excitation temperature and velocity dispersion. In Section 4, we present the results: the presence of multiple components, the derived deuteration ratios, and the turbulence on small scales. In Section 5, we discuss and analyse the results: showing the effects of the filling factor on the deuteration ratio, the presence and influence of a 70 $\upmu$m source in the region, and we provide an estimate to the age of the regions. Finally in Section 6, we summarize our results.

\section{Data}

\subsection{Catalogue and initial MOPRA observations}

The Spitzer Dark Cloud (SDC) catalogue is a survey and analysis of IRDCs in the Galaxy using the \textit{Spitzer} GLIMPSE 8$\upmu$m data (Peretto \& Fuller, 2009). More than 11000 sources are identified in $10\degr < |l|< 65\degr$ and $0\degr < |b| < 1\degr$. From this catalogue we picked individual IRDCs or IRDCs-complexes to analyse the dense gas distribution and properties. Individual IRDCs were selected because they have a filamentary shape. The IRDCs-complexes were selected because of their filamentary shape in a larger scale.

Using the Australia Telescope National Facility (ATNF) MOPRA 22m telescope we observed molecular lines at 3mm, including the HNC (1--0) and N$_2$H$^+$ (1--0) lines, towards these objects. The observations were carried out in 2010 May, 2011 July and 2012 July.  The data was reduced using \textsc{gridzilla} \footnote{http://www.atnf.csiro.au/computing/software/livedata/index.html}. The MOPRA beam FWHM is 36 arcsec at the observed wavelength. The noise rms range was 60--330 mK after smoothing the spectra to a 0.22 km s$^{-1}$ channel spacing. One long IRDCs-complex was mapped, and eight individual IRDCs were mapped. Additionally 66 single pointings were made towards the densest parts of 14 IRDCs. The large IRDC complex, which we call L332, is a $\sim 0.5\degr$ long which encompasses more than 36 individual IRDCs. The HNC and N$_2$H$^+$ (1--0) maps exhibits a coherent structure connecting all the single IRDCs (Lackington et al., in preparation).

The N$_2$H$^+$ (1--0) spectra were reduced using the procedure described in Section 3, and the results of fitting the hyperfine structure are shown in Table \ref{tab:pars_n2h10}. 

\subsection{APEX}

\begin{figure*}
\includegraphics[width=0.7\textwidth]{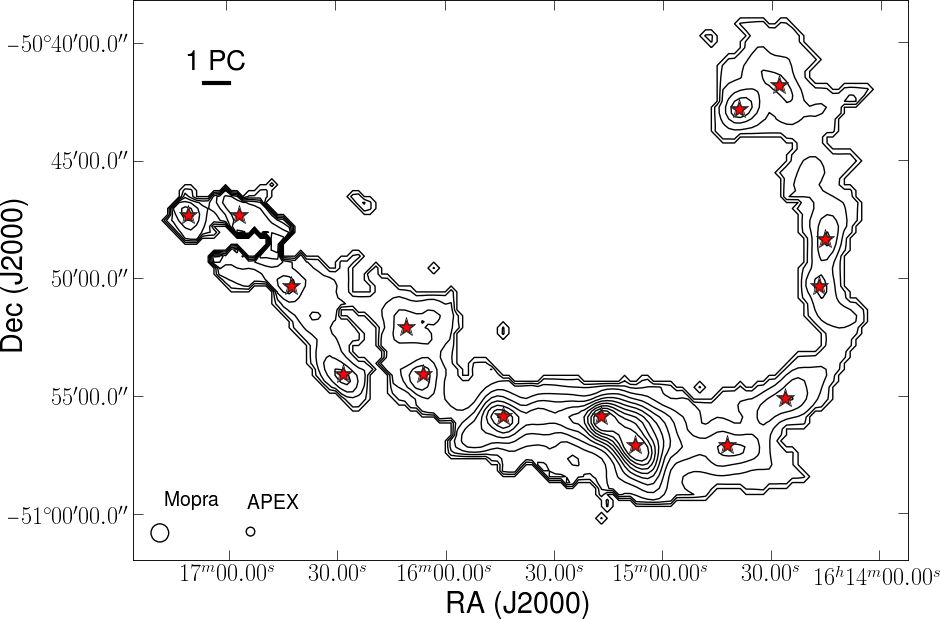}
\caption[MOPRA HNC (1--0) velocity integrated intensity map]{MOPRA HNC (1--0) velocity integrated intensity map towards the filamentary complex of IRDCs, L332. Contours represent 90, 80,. . ., 20, 10, 5 per cent of the peak value. The red stars represent the selected positions to observe N$_2$H$^+$ and N$_2$D$^+$ (3--2) using APEX. In lower-left, we show the MOPRA and APEX beams for comparison. }
\label{fig:l332apex}
\end{figure*}

From the MOPRA maps and single-pointing spectra we selected the brightest cores in these IRDCs. Since the HNC(1--0) and N$_2$H$^+$ (1--0) peaks are well correlated (Lackington et al., in preparation) and HNC has higher signal to noise we used the peaks identified by the HNC (1--0) emission. Fig.~\ref{fig:l332apex} shows how these positions were selected in the L332 IRDC-complex. Using the Atacama Pathfinder EXperiment (APEX) telescope we observed N$_2$H$^+$ (3--2) and N$_2$D$^+$ (3--2) towards the selected cores. For the selected lines the APEX beam is smaller (22--27 arcsec) than the MOPRA beam (Fig.~\ref{fig:l332apex}), meaning we are probing smaller scales. The observations of N$_2$H$^+$ and N$_2$D$^+$ (3--2) were carried out in 2012 August--November with APEX. In total 54 N$_2$H$^+$ (3--2) single-pointing observations (Table \ref{tab:nobs}), using the APEX-2 receiver of the Swedish Heterodyne Facility Instrument \citep[SHeFi;][]{Vassilev2008}, were made. The cores with the brightest and narrowest N$_2$H$^+$ (3--2) line were also observed in N$_2$D$^+$ (3--2) using the APEX-1 receiver of the SHeFi. In total 29 cores were observed in N$_2$D$^+$ (3--2). That makes this list the largest sample of deuterium observations towards high-mass star-forming regions. The positions observed in N$_2$H$^+$ and N$_2$D$^+$ (3--2) are listed in Table \ref{tab:sources}. Instead of using individual IRDCs names to identify the positions observed in the L332 complex we numbered the 15 peaks observed in the region, and we use this as the identifier.

\begin{table}
\caption{Summary of observations for observed transitions.}
\begin{tabular}{|l|r|r|}
\hline
Transition & \multicolumn{1}{l|}{Objects observed} & \multicolumn{1}{l|}{Objects detected} \\ \hline \hline
N$_2$H$^+$ (3--2) & 53 & 53 \\
N$_2$H$^+$ (3--2) & 54 & 51 \\
N$_2$D$^+$ (3--2) & 29 & 13 \\ \hline
\end{tabular}
\label{tab:nobs}
\end{table}

The observations used position switching, and were calibrated by regularly measuring the sky brightness and cold/hot loads in the cabin. All observations used the extended bandwidth Fast Fourier Transform Spectrometer (XFFTS), which provides two 2.5 GHz units, with 32 768 channels each unit and 1 GHz overlap. This configuration provides an spectral resolution of 0.082 and 0.099 km s$^{-1}$ for N$_2$H$^+$ (3--2) and N$_2$D$^+$ (3--2), respectively. The N$_2$H$^+$ (3--2) observations had a T$_{sys}$ range of 117--150 K, and noise rms range of 58--400 mK. For the N$_2$D$^+$ (3--2) observations the T$_{sys}$ range was 145--158 K, and the noise rms range was 24--67 mK after smoothing the spectra to a 0.197 km s$^{-1}$ channel spacing. The APEX half power beam width (HPBW) for N$_2$H$^+$ (3--2) is 22 arcsec, which at a typical distance for the sample of 3.1 kpc (see Section 4.3) corresponds to a typical size of 0.33 pc. On the other hand the APEX HPBW for N$_2$D$^+$ (3--2) is 27 arcsec which means a size of 0.4 pc at 3.1 kpc.

All spectra were reduced using \textsc{class}90. Before averaging all the scans, each spectrum had the baseline removed by fitting a first- or second-order polynomial.

\begin{table}
\caption{Observed positions. The columns show the source name, right ascension (J2000), declination (J2000), and number of young stellar objects (YSO) within the APEX beam for N$_2$D$^+$ (3--2). }
\begin{tabular}{lllr}
\hline \hline
Source & RA (J2000) & Dec. (J2000) & \multicolumn{1}{l}{YSOs in beam} \\ 
\hline
SDC299.2 & 12:18:26.289 & -63:01:31.710 & 0 \\ 
SDC320.271 & 15:07:56.547 & -57:54:16.420 & 1 \\ 
SDC320.252 & 15:07:21.101 & -57:49:14.420 & 1 \\ 
SDC321.708 & 15:18:01.201 & -57:21:55.150 & 1 \\ 
SDC321.758 & 15:18:26.711 & -57:21:44.740 & 0 \\ 
SDC321.936 & 15:19:42.442 & -57:17:50.750 & 0 \\ 
SDC329.028 & 16:00:30.790 & -53:12:29.200 & 1 \\ 
SDC329.068 & 16:01:08.860 & -53:15:38.500 & 0 \\ 
SDC329.186 & 16:01:46.800 & -53:11:45.000 & 1 \\ 
SDC329.313 & 16:02:23.440 & -53:05:59.700 & 1 \\ 
SDC351.438 & 17:20:53.900 & -35:45:21.000 & 0 \\ 
SDC352.027 & 17:22:33.937 & -35:13:11.220 & 0 \\ 
SDC351.8 & 17:21:49.399 & -35:27:31.150 & 0 \\ 
SDC351.562 & 17:20:52.217 & -35:35:09.840 & 0 \\ 
L332-10 & 16:14:42.115 & -50:57:07.776 & 0 \\ 
L332-11 & 16:14:26.280 & -50:55:07.356 & 0 \\ 
L332-12 & 16:14:16.874 & -50:50:22.056 & 0 \\ 
L332-13 & 16:14:15.336 & -50:48:21.996 & 0 \\ 
L332-14 & 16:14:28.121 & -50:41:52.404 & 0 \\ 
L332-15 & 16:14:39.156 & -50:42:52.704 & 0 \\ 
L332-1 & 16:17:10.946 & -50:47:20.004 & 0 \\ 
L332-3 & 16:16:42.559 & -50:50:21.444 & 1 \\ 
L332-4 & 16:16:28.397 & -50:54:06.984 & 1 \\ 
L332-5 & 16:16:10.920 & -50:52:07.536 & 0 \\ 
L332-6 & 16:16:06.197 & -50:54:07.668 & 0 \\ 
L332-7 & 16:15:44.011 & -50:55:53.040 & 1 \\ 
L332-8 & 16:15:17.038 & -50:55:53.184 & 1 \\ 
L332-9 & 16:15:07.512 & -50:57:08.136 & 0 \\ 
SDC335.283A & 16:29:01.428 & -48:50:28.464 & 1 \\ 
SDC335.283B & 16:28:50.800 & -48:50:42.300 & 0 \\ 
SDC335.077A & 16:29:22.270 & -49:12:14.616 & 0 \\ 
SDC335.077B & 16:29:28.900 & -49:10:54.500 & 0 \\ 
SDC335.579A & 16:31:01.900 & -48:44:04.000 & 0 \\ 
SDC335.579B & 16:30:54.800 & -48:42:43.000 & 0 \\ 
SDC335.579Y1 & 16:30:58.800 & -48:43:54.000 & 1 \\ 
SDC335.579Y2 & 16:30:57.300 & -48:43:40.000 & 1 \\ 
SDC335.44 & 16:30:03.900 & -48:48:38.000 & 0 \\ 
SDC335.253 & 16:29:42.900 & -48:58:48.000 & 0 \\ 
SDC335.229A & 16:29:35.800 & -48:59:48.000 & 0 \\ 
SDC335.229B & 16:29:38.800 & -49:00:40.000 & 0 \\ 
SDC335.059 & 16:29:11.700 & -49:11:36.000 & 0 \\ 
SDC356.84 & 17:37:57.149 & -31:36:13.176 & 0 \\ 
SDC13.174A & 18:14:30.413 & -17:33:08.712 & 1 \\ 
SDC13.174B & 18:14:26.220 & -17:31:53.688 & 0 \\ 
SDC13.194 & 18:14:33.559 & -17:31:08.688 & 0 \\ 
SDC16.915 & 18:21:55.877 & -14:15:03.672 & 1 \\ 
SDC2.141 & 17:49:32.652 & -26:57:07.812 & 0 \\ 
SDC2.116 & 17:49:24.283 & -26:58:24.060 & 0 \\ 
SDC301.876 & 12:42:19.298 & -62:14:40.200 & 1 \\ 
SDC327.894 & 15:51:52.740 & -53:26:11.688 & 1 \\ 
SDC327.964 & 15:52:31.486 & -53:26:50.388 & 0 \\ 
SDC34.558 & 18:52:34.358 & +01:41:11.580 & 0 \\ 
SDC37.08 & 18:59:04.134 & +03:38:31.670 & 1 \\ 
\hline
\end{tabular}
\\
\begin{flushleft}
\end{flushleft}
\label{tab:sources}
\end{table}

%section 3
\section{Analysis}

The N$_2$H$^+$ (1--0), N$_2$H$^+$ (3--2) and N$_2$D$^+$ (3--2) transitions have hyperfine structure, which allows independent calculation of the optical depth, $\tau_0$, and the excitation temperature, \tex. The lines were fitted in \textsc{python} using the Levenberg--Marquardt least-squares minimization routine \emph{mpfit}. The model we used takes into account all the hyperfine components, and depends on four variables: optical depth ($\tau_0$), excitation temperature (\tex), central velocity ($v_{\text{c}}$), velocity dispersion ($\sigma$; also described as the linewidth hereafter).  The approach we use is similar to the one by \citet{Fuller1993}, \citet{Rosolowsky2008} and \citet{Pineda2013}. The expected emission is,

\begin{equation}
\label{eq:hfs1}
T_{\text{mb}}(\nu) = A[1 - e^{-\tau(\nu)}],
\end{equation}
where T$_{\text{mb}}$ is the main beam temperature, which was obtained using the telescope efficiency, 0.74 for N$_2$H$^+$ (3--2), and 0.75 for N$_2$D$^+$ (3--2). Also

\begin{equation}
\label{eq:hfs2}
\tau(\nu) = \tau_0\,\sum_{i=1}^{N} a_i\,e^{-[(v-v_{\text{c}})\,+\,v_i]^2/2\sigma^2},
\end{equation}

\begin{equation}
\label{eq:hfs_norm}
\sum_{i=1}^{N} a_i = 1,
\end{equation}

\begin{equation}
\label{eq:hfs3}
A = \eta_{\text{f}} [J(T_{\text{ex}}) - J(T_{\text{bg}})],
\end{equation}
and,

\begin{equation}
\label{eq:hfs4}
J(T) = {h\nu \over k}{1 \over \exp(h\nu/kT)-1},
\end{equation}
where $v_i$ is the velocity offset of the ith hyperfine component, and $a_i$ is its statistical weight, and $\eta_{\text{f}}$ is the beam filling factor.

N$_2$H$^+$ (3--2) has 45 hyperfine components, and N$_2$D$^+$ (3--2) has 25 hyperfine components. The rest frequencies used were 231.3219 and 279.5118 GHz for N$_2$D$^+$ (3--2) and N$_2$H$^+$ (3--2) respectively. Due to the linewidths of our sample the N$_2$H$^+$ (3--2) hyperfine components are blended and exhibit at most three blended components. On the other hand the linewidths and noise of N$_2$D$^+$ (3--2) observations blend its hyperfine components into only one peak feature above the noise level.

Since the spatial extent of the emission is unknown we assume $\eta_{\text{f}} = 1,$ which provides us with a lower limit to the true column densities. With this assumption \tex\ is obtained from the fit when $\tau_{0}$ is well constrained. In addition, $\tau_{0}$ is limited in the 0.1--30 range as outside this limits the equation becomes degenerate. In the optically thick limit exp($-\tau_{0}) \ll$ 1, and in the optically thin limit \mbox{1 -- exp($-\tau_{0}) \approx \tau_{0}$} making $\tau_{0}$ and \tex\ degenerate. To avoid the optically thin degeneracy if $\tau_{0} \leq 0.1$, and thus \tex\ is not constrained, we assumed $\tau_{0} = 0.1$. Using this criteria, three sources are optically thin.

\begin{figure}
\includegraphics[width=0.45\textwidth]{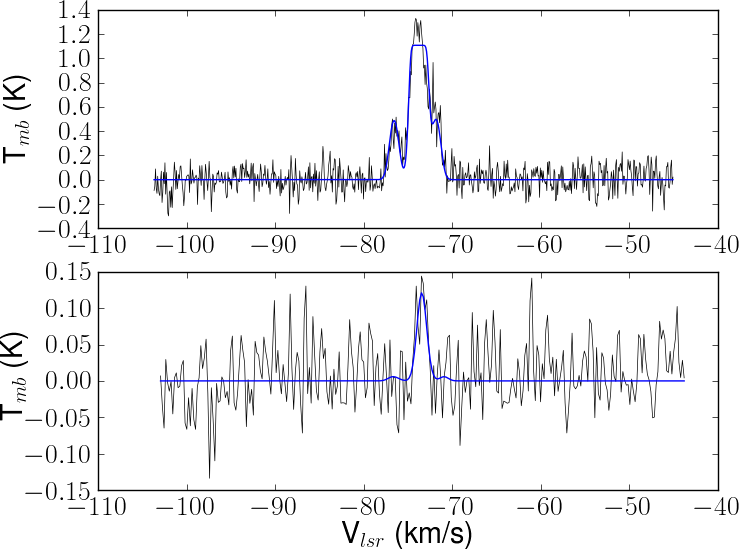}
\caption[]{SDC329.313 spectra: N$_2$H$^+$ (3--2) (top) and N$_2$D$^+$ (3--2) (bottom) sample spectra overplotted with the hyperfine component model fit.}
\label{fig:spectra}
\end{figure}

Example spectra N$_2$H$^+$ (3--2) and N$_2$D$^+$ (3--2) are shown in Figure \ref{fig:spectra}. Both lines towards this object are well fitted using a single velocity component which is shown in the figure. Similar figures for all the sources are shown in the appendix which is available online.

\subsection{Distances}

Distances to the clouds were computed using an adaptation to \textsc{python} from the \textsc{idl} routine 'kindist.pro', which uses a galactic rotation curve to calculate kinematic distances. The constants used are 8.4 kpc of galactocentric distance and local standard of rest (LSR) velocity of 254 km s$^{-1}$ around the galactic centre \citep{Reid2009}. For all sources except for the group around of four pointings around SDC351.8 we obtained reasonable values. The calculated values for these sources varied from 250pc to 1kpc, a spread unlikely given that they are from the same region and have similar velocities. In addition, 250 pc is very small compared to the rest of the sample which also seems unlikely. So for this source we will use the mean distance of the sample, 3.1 kpc. For SDC299.2 and SDC301.876 the measured central velocity was beyond the tangent point value, so again we used the mean distance.

%section 4
\section{Results}

The N$_2$H$^+$ (3--2) spectra have typical noise levels of 0.124 K in the 0.08 \kms\ velocity channels while the N$_2$D$^+$ (3--2) spectra has typical noise levels of 0.04 K in 0.2 \kms\ velocity channels. Meanwhile the N$_2$H$^+$ (1--0) spectra have typical noise levels of 0.11 K after smoothing the spectra to a 0.22 \kms\ velocity channels. We detected N$_2$H$^+$ (3--2) emission towards 51 of the 54 pointings. Only SDC13.174B, SDC335.253, and SDC335.077B did not exhibit N$_2$H$^+$ (3--2) emission. N$_2$D$^+$ (3--2) emission was detected towards 13 of the 29 pointings (Table \ref{tab:nobs}). Finally N$_2$H$^+$ (1--0) was detected towards 52 of the 53 pointings.

For 38 out of the 51 N$_2$H$^+$ (3--2) observations a good fit is obtained with a single velocity component. Tables \ref{tab:pars_n2h} and \ref{tab:pars_n2h_p2} shows the results of the hyperfine structure fitting to the N$_2$H$^+$ (3--2) line of the detected objects. A two-component fit is necessary in 13 cases, which present two distinct components along the line of sight or line asymmetry consistent with an additional component along the line of sight. In these cases two independent velocity components are fitted simultaneously, with each component labelled `-a' or `-b'. The fitted parameters for the two velocity components are shown in Tables \ref{tab:pars_n2h} and \ref{tab:pars_n2h_p2}. On the other hand all the N$_2$H$^+$ (1--0) detections are well fitted with only one velocity component.

For N$_2$H$^+$ (3--2) the fitting procedure gives well-constrained parameters for all except 15 sources. These 15 sources had poorly constrained optical depth ($\tau_0 < 3\sigma_{\tau}$, where $\sigma_{\tau}$ is the estimated uncertainty in the optical depth). For these sources we fix the excitation temperature to the average value obtained from all other spectra, i.e., 6.5 K. This removes one free parameter in the fit providing better constrained estimates of the other parameters.

Due to the low signal to noise and blending of the hyperfine components of the N$_2$D$^+$ (3--2), the full four-parameter fit gives badly constrained parameters. Since both transitions have similar critical densities, we fix the N$_2$D$^+$ (3--2) excitation temperature to the value from the N$_2$H$^+$ (3--2) fit. This procedure is similar to the one followed by \citet{Crapsi2005} and \citet{Pineda2013}. For N$_2$D$^+$ (3--2), in all cases the optical depth obtained from the hyperfine structure fit is lower than 1. Therefore, we assume that N$_2$D$^+$ (3--2) emission is optically thin in the further analysis. Table \ref{tab:pars_n2d} shows the results of the hyperfine structure fitting to the N$_2$D$^+$ (3--2) line.

N$_2$D$^+$ (3--2) was detected towards 13 of the 29 pointings. The detections may be because there is something inherently different in these objects. To evaluate this possibility we separated the two samples, detections and non-detections of N$_2$D$^+$ (3--2), and performed statistical tests on the derived properties. We use the Kolmogorov--Smirnov test to determine if the derived properties of both samples follow the same distribution, and Student's t-tests to evaluate if the samples means are statistically different. We compare the detected and non-detected samples on the N$_2$H$^+$ (3--2) velocity dispersion, integrated intensity, column density, optical depth, and excitation temperature.

\begin{figure}
\includegraphics[width=0.47\textwidth]{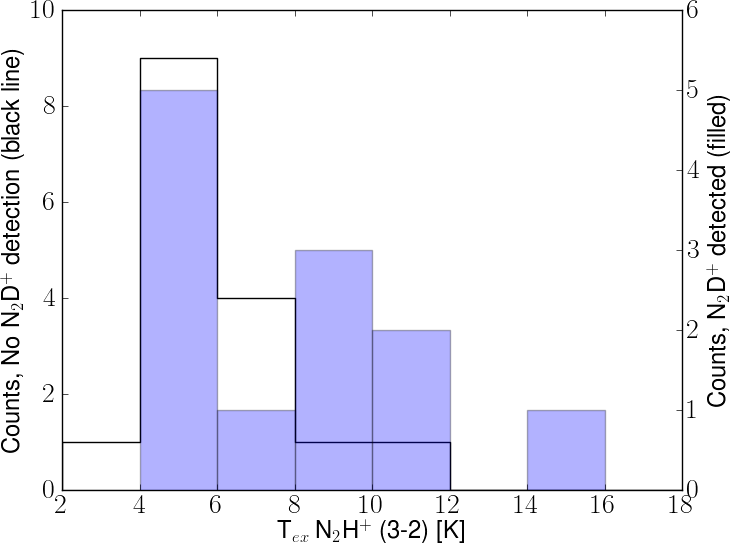}
\caption[]{Histogram of N$_2$H$^+$ (3--2) excitation temperature for two subsamples of the sources with observations of N$_2$H$^+$ (3--2): filled is the histogram for sources with detected N$_2$D$^+$ (3--2) (right hand scale), and the black line is the histogram for sources without detection (left-hand scale). }
\label{fig:histo_tex}
\end{figure}

The tests show no significant difference between the detected and non-detected objects except in the excitation temperature, \tex. For all the other properties we can not reject the null hypothesis, of identical distributions, at more than a 0.05 level. The distributions of \tex\ for the detected and non-detected sources are shown in Fig. \ref{fig:histo_tex}. The detected sources have a mean \tex\ of 9.0 $\pm$ 0.4, while the non-detections appear cooler with a mean \tex\ of 6.2 $\pm$ 0.1, perhaps suggesting a difference in density between the objects. However, whether the non-detections reflect the intrinsic difference in the abundance of N$_2$D$^+$, or the excitation of N$_2$D$^+$ and the detectability of the 3-2 transitions awaits a detailed analysis of the N$_2$D$^+$ excitation in these sources.

We calculated the column density of N$_2$H$^+$ (1--0), N$_2$H$^+$ (3--2) and N$_2$D$^+$ (3--2) using equations in appendix A of \citet{Pineda2013}. For both N$_2$H$^+$ transitions we used the optically thin case if $\tau_0 < 0.1$ or $\tau_0 < 3 \sigma_{\tau}$ (three sources). For N$_2$D$^+$ (3--2) we use the optically thin equation. Assuming a filling factor of unity, the derived column densities represent a lower limit of the true values.

The calculated N$_2$H$^+$ column density varies depending on which transition is used. For sources with relatively low (1--0) derived column densities but apparently large ($>$ 5) (3--2) optical depths, the (3--2) derived column densities are much larger (about 10--80 times) than the (1--0) derived column densities. We attribute this to two factors: overestimation of the optical depth of the (3--2) transition (and underestimation of its uncertainties) in part due to the interdependence of the optical depth and velocity dispersion when fitting the lines, and difference in filling factors in both transitions. Because of the possible overestimation of the (3--2) optical depth, the (1--0) derived column densities should be more reliable.

\subsection{Line modelling}

We attempted to correct the incongruence of the different derived column densities by modelling the N$_2$H$^+$ emission using a non-LTE molecular radiative transfer, \textsc{radex} \citep{vanderTak2007}\footnote{http://www.sron.rug.nl/$\sim$vdtak/radex/radex.php}. For each object we run \textsc{radex} using as inputs: the observed N$_2$H$^+$ (1--0) linewidth, a range of kinetic temperatures (5--20 K), a range of H$_2$ densities (10$^5$--2$\times$10$^6$ cm$^{-3}$), and a range of N$_2$H$^+$ column densities (0.1--10 times the N$_2$H$^+$ (1--0) derived value). In addition we introduced a filling factor for the (1--0) transition in the analysis of the models.

To discriminate between all the models we used the observed integrated intensity which does not depend on any modelling or line fitting, and the optical depth of the (1--0) which can be obtained with high certainty as it is well constrained by the hyperfine fit. We produced a first cut by selecting the \textsc{radex}-derived models that had an N$_2$H$^+$ (1--0) optical depth equal within 10\% of the observed value or within the error in the observation, whichever was greater. From the remaining models we produced a second cut by only keeping only the models with N$_2$H$^+$ (1--0) integrated intensity within 10\% of the observed value or within the error in the observation. To do this a beam filling factor was introduced for the models. Finally only the models where the N$_2$H$^+$ (3--2) integrated intensity matched the observation (again within 10\% or the observation error) were selected.

\begin{figure}
\includegraphics[width=0.47\textwidth]{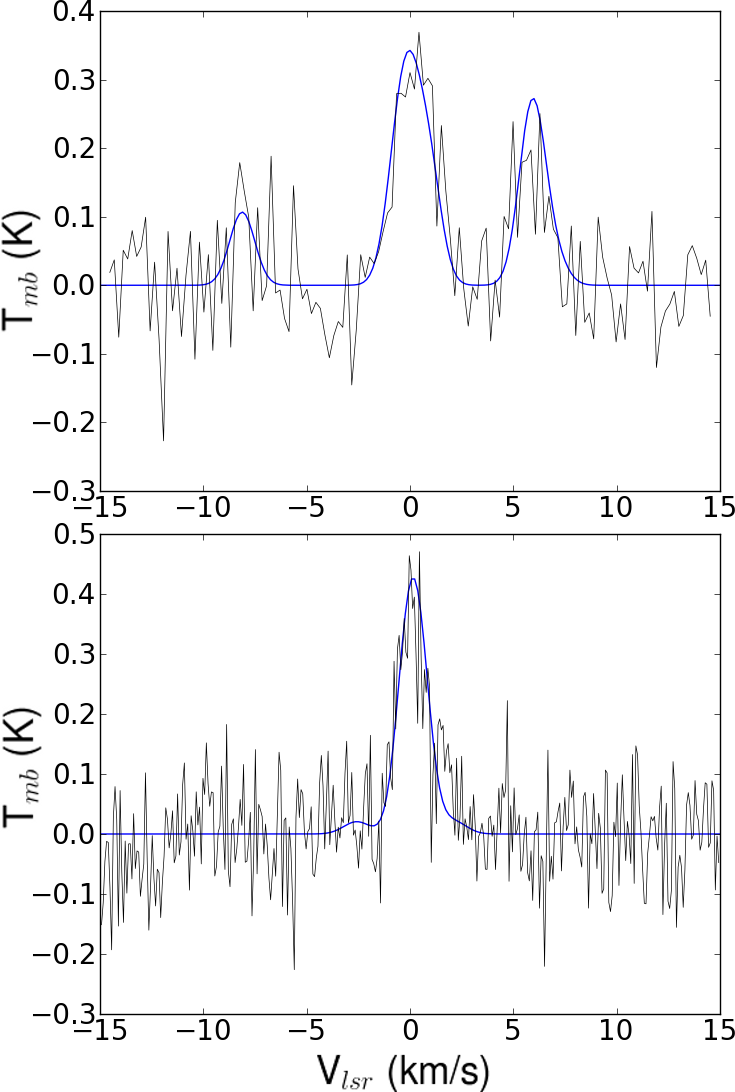}
\caption[]{Example of \textsc{radex} derived model for N$_2$H$^+$ towards L332-11. Top: \textsc{radex} model (blue line) over the observed N$_2$H$^+$ (1--0) spectra. Bottom: the same \textsc{radex} model over the observed N$_2$H$^+$ (3--2) spectra. In both cases, the obtained \textsc{radex} model fits the observed spectra very well.}
\label{fig:radex_model}
\end{figure}

\begin{figure}
\includegraphics[width=0.47\textwidth]{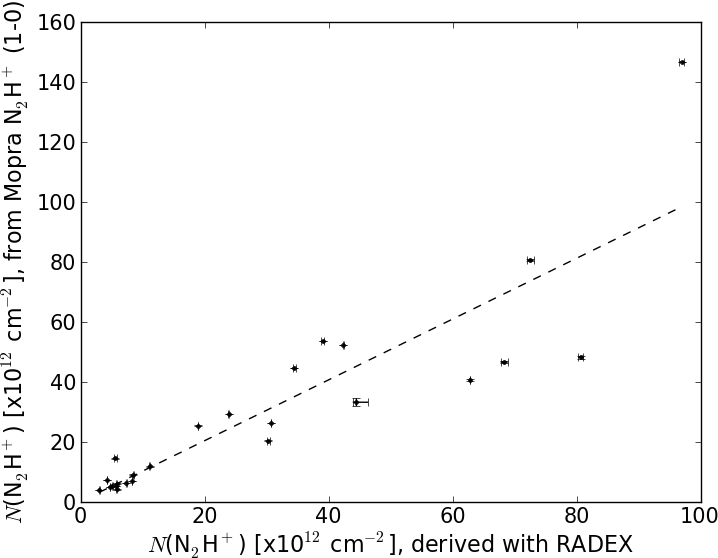}
\caption[]{\textsc{radex} derived N$_2$H$^+$ column density versus N$_2$H$^+$ column density derived from the MOPRA N$_2$H$^+$ (1--0) observations, using a 0.4 filling factor for both values. The dashed line represent the linear regression of the values. The slope of the linear regression is 1.01 meaning that the derived models on average represent a good approximation of the observed values.}
\label{fig:radex_cdens}
\end{figure}

The number of objects for which models reproduce the observations varied for each filling factor. The best results were obtained with a filling factor of 0.4 for the N$_2$H$^+$ (1--0) transition, where half of the objects had models (more than for any other filling factor used). Fig. \ref{fig:radex_model} shows an example of how the final \textsc{radex} model fit the observed N$_2$H$^+$ (1--0) and (3--2) lines. Assuming equal sizes of the emission regions for (1--0) and (3--2) transitions, a beam filling factor of 0.4 for the (1--0) emission implies a filling factor of unity for the (3--2) beam. Fig. \ref{fig:radex_cdens} shows the result of the modelling. The mean value of the N$_2$H$^+$ column density for all the models obtained of each object is plotted versus the N$_2$H$^+$ column density derived from the (1--0) observation using a filling factor of 0.4. It is remarkable that there is a very good agreement between the two values. This is quantified by calculating the linear regression of the values. The obtained slope 1.01 and the r-value is 0.88, meaning that on average the models represent a good fit of the observations.

The kinetic temperatures obtained through the \textsc{radex} modelling range from 5.6 to 19.1 K. The \textsc{radex} obtained densities vary from 2.9 $\times$ 10$^{5}$ to 2 $\times$ 10$^{6}$ cm$^{-3}$.

Thus a filling factor of 0.4 is a good approximation for this sample. So for the rest of the paper we use the N$_2$H$^+$ column density derived from the (1--0) observations using a filling factor of 0.4. Table \ref{tab:pars_n2h10} lists this derived value. Tables \ref{tab:pars_n2h} and \ref{tab:pars_n2h_p2} list the value of the (3--2) derived N$_2$H$^+$ column density for reference. Finally Table \ref{tab:pars_n2d} lists the N$_2$D$^+$ column density which assumes a filling factor of unity.

The modelling implies a fractional abundance of N$_2$H$^+$ of 3.6 $\times 10^{-10}$ to 2.6 $\times 10^{-8}$ for the different objects. These values are similar to those observed by other studies. Using MOPRA \cite{Miettinen2014} found fractional abundances of 2.8 $\times 10^{-10}$--9.8 $\times 10^{-9}$ towards IRDCs. Similarly \cite{Sanhueza2012} found abundances of 1.9 $\times 10^{-10}$--1.67 $\times 10^{-8}$.

\subsection{Multiple Components}

\begin{figure}
\includegraphics[width=0.47\textwidth]{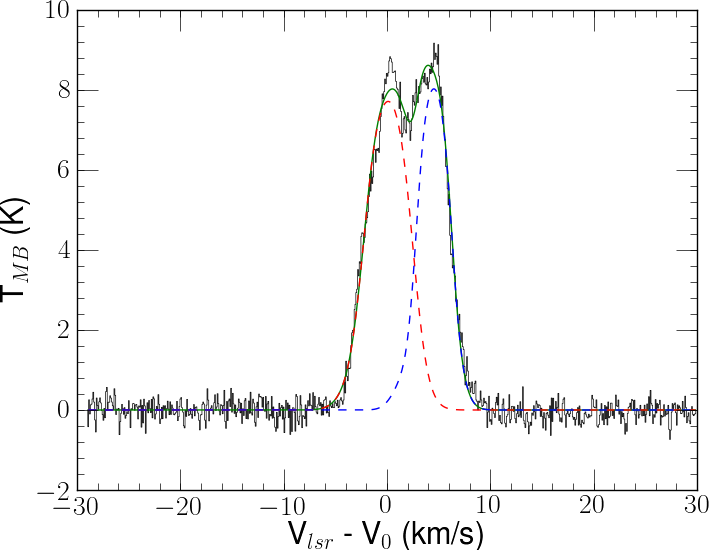}
\includegraphics[width=0.47\textwidth]{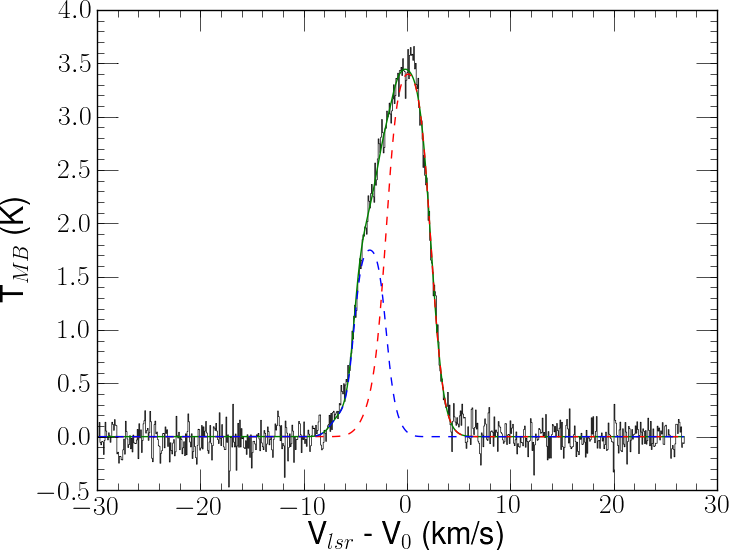}
\caption[]{Two examples of N$_2$H$^+$ (3--2) spectra, showing evidence of two velocity components. Both images show the N$_2$H$^+$ (3--2) spectra overplotted with the two hyperfine component model fits (red and blue dashed lines) and the total model fit (green line). Top: SDC351.438, the two components seen clearly as a two-peaked feature. Bottom: L332-8, the two components are necessary to explain the broad feature of the N$_2$H$^+$ (3--2) line. }
\label{fig:2c_example}
\end{figure}

Fig. \ref{fig:2c_example} shows spectra of N$_2$H$^+$ (3--2) towards two sources, which exhibit two velocity components. These spectra exemplify the two types of two-component spectra in the sample: Fig. \ref{fig:2c_example} top (SDC351.438) shows a two-peaked feature, and Fig. \ref{fig:2c_example} bottom (L332-8) shows a very broad feature which is likely due to two blended components. Two component features are seen in the N$_2$H$^+$ (3--2) towards 13 of the 51 sources (25\%).

There are several explanations for these two component lines. Optical depth effects could explain some of the sources. The more likely explanation however is the presence of multiple velocity components along the line of sight, as seen in high-mass star formation simulations \citep{Smith2013} and in a number of observations \citep{Csengeri2011,Beuther2013,Peretto2013}.

There is no evidence to suggest that the presence of deuteration is correlated with the presence of multiple velocity components.
Of the 28 objects with N$_2$H$^+$ and N$_2$D$^+$ (3--2) observations, eleven had two N$_2$H$^+$ (3--2) components, ie (39\%). Of the sources detected in N$_2$D$^+$ (3--2) six of thirteen had two-velocity components, ie 46\%. Chance of randomly selecting six or more two components out of 13 picks is 38\%, consistent with a random selection.

In the remainder of the analysis, we associate the velocity component of N$_2$D$^+$ with the NN$_2$H$^+$ component closest in velocity.

\subsection{Deuteration Ratio}

The deuteration ratio, D$_\textrm{R}$, is defined as the ratio of the column density of a deuterated species to its undeuterated counterpart. So for N$_2$H$^+$,
\begin{equation}
\label{eq:dr}
\rmn{D}_R =N \rmn{(N_2 D^+) /} N \rmn{(N_2 H^+) }.
\end{equation}

\begin{figure}
\includegraphics[width=0.47\textwidth]{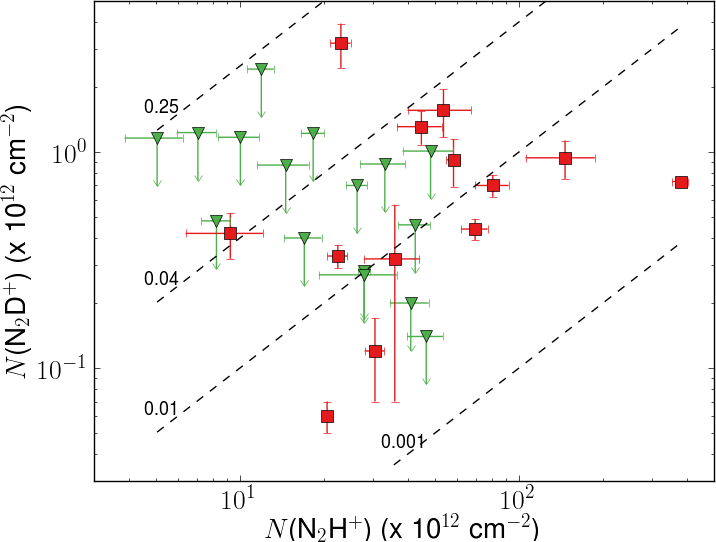}
\caption[]{N$_2$H$^+$ column density versus N$_2$D$^+$ column density. Dashed lines are constant D$_\textrm{R}$ ratio of 0.01, 0.04, and 0.25. Red squares represent detections, and green triangles represent upper limits of the N$_2$D$^+$ column density. }
\label{fig:ratio_ul}
\end{figure}

Figure \ref{fig:ratio_ul} shows the column density of N$_2$D$^+$ versus the column density of N$_2$H$^+$, overlaid with lines of constant D$_\textrm{R}$ = 0.01, 0.04, and 0.25. Over the sample of objects with detected N$_2$D$^+$, D$_\textrm{R}$ has a mean of 0.024, a standard deviation of 0.037, and reaches a maximum value of 0.14. As we are using a filling factor of unity for N$_2$D$^+$, the N$_2$D$^+$ column density is a lower limit of the true value, thus the observed values of D$_\textrm{R}$ is also a lower limit value. On the other hand the real average ratio may be lower because of the non-detections and upper limits in the sample.

\subsubsection{Comparison with previous observations}

\citet{Crapsi2005} did a survey of 31 low-mass starless cores in N$_2$H$^+$ (1--0), N$_2$D$^+$ (2--1), 13 of which had C$^{18}$O, and 1.2mm continuum observations. Their main objective was to investigate the evolution of the cores. For this they used several methods, such as determining the abundance of N$_2$H$^+$ and N$_2$D$^+$, deuteration ratios, and CO depletion factors. For the mapped cores N$_2$H$^+$ and N$_2$D$^+$ peak close to the dust continuum maxima.

\citet{Emprechtinger2009} observed 20 class 0 protostars using the IRAM-30m telescope in various lines including N$_2$H$^+$ (1--0) and N$_2$D$^+$ (2--1). They determined the deuterium fractionation in order to analyse the chemical evolution of the objects. Compared with pre-stellar cores their class 0 objects had wider linewidths, higher excitation temperature and lower optical depth, consistent with the presence of a protostar. They found that D$_\textrm{R}$ correlates with dust temperature, CO depletion factor, bolometric luminosity. This result is consistent with models where the maximum D$_\textrm{R}$ is at the moment of collapse, and decreases as the heating of the protostar unfreezes the CO of the inner parts.

\citet{Fontani2011} took observations of 27 objects at different evolutionary stages of high-mass star formation process in N$_2$H$^+$(3--2) and N$_2$D$^+$(2--1). They studied high-mass starless cores candidates (HMSCs, 10 sources), high-mass protostellar objects (HMPOs, 10 objects), and ultra compact H\textsc{ii} regions (UCH\textsc{ii}s, 7 objects). Their objective was to determine if the deuteration ratio is an evolutionary tracer in high mass regions as in low-mass regions. They found that the abundance of N$_2$D$^+$ is higher in HMSCs than in HMPOs and UCH\textsc{ii} regions, and that objects with highest deuteration ratio are starless cores. Their results indicate that D$_R$ can be used as an evolutionary tracer in massive star forming regions, and that the physical conditions acting on the abundance of deuterated species likely evolve similarly in the low and high mass star formation processes.

Observations of the Spokes Cluster or NGC2264-D show supersonic turbulence in N$_2$H$^+$ (1--0) \citep{Peretto2006}, however the fragmentation of most massive cores suggests that turbulence was not dominant. \citet{Pineda2013} set out to determine if high density gas in NGC2264-D is subsonic or less turbulent, which would reconcile previous observations. They observed 14 cores in N$_2$H$^+$ (3--2) and eight of them in N$_2$D$^+$ (3--2) using APEX, in order to study the line widths and deuteration ratios. The N$_2$H$^+$ and N$_2$D$^+$ (3--2) show narrower line widths than N$_2$H$^+$ (1--0). Three cores with no protostar show the highest level of deuteration, and are likely the youngest. They conclude that N$_2$H$^+$ and N$_2$D$^+$ (3--2) probe more quiescent gas and therefore is a better tracer of this gas in high-mass regions.

\begin{figure}
\includegraphics[width=0.47\textwidth]{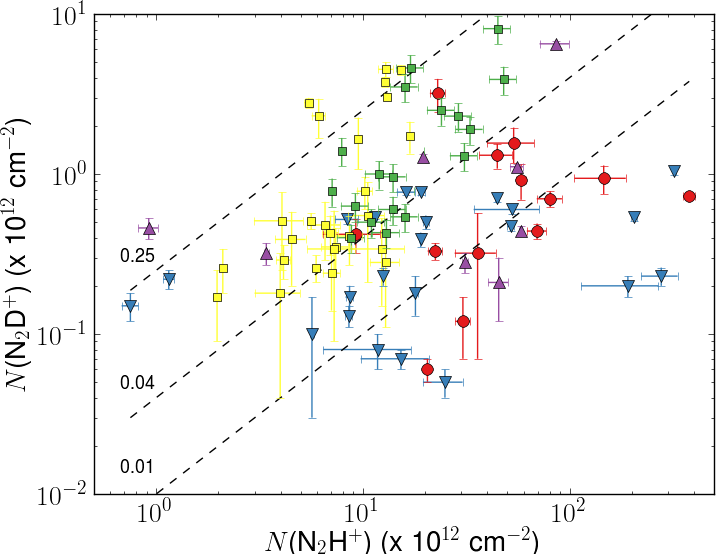}
\caption[]{N$_2$H$^+$ column density versus N$_2$D$^+$ column density. Dashed lines are constant D$_\textrm{R}$ ratio of 0.01, 0.04, and 0.25. Red circles = this paper. Purple upward triangles = \citet{Pineda2013}. Blue downward triangles = \citet{Fontani2011} (Recalculated values, upper limits not included). Yellow squares = \citet{Crapsi2005}. Green squares = \citet{Emprechtinger2009}.}
\label{fig:ratio_final}
\end{figure}

In Fig. \ref{fig:ratio_final} we compare our results and the results from \citet{Crapsi2005}, \citet{Emprechtinger2009}, \citet{Fontani2011},  and \citet{Pineda2013}. The values of the column densities of \citet{Fontani2011} were recalculated using our procedure assuming equal filling factors, and we did not show objects that have only upper limits in N$_2$D$^+$.

As in Section 4 we use KS-tests and Student's t-tests to investigate whether these samples have statistically different distributions. As expected, all except one of our regions have N$_2$H$^+$ column density greater than all the low-mass sources of \citet{Crapsi2005}, confirmed with the tests (p-values $<$ 0.01). In contrast the N$_2$D$^+$ column densities are similar in both samples (p-values $>$ 0.1, we can not reject the null hypothesis of identical samples), which leads to the conclusion that our regions tend to have smaller deuteration ratios. On the other hand we are observing about the same range of N$_2$H$^+$ and N$_2$D$^+$ column densities as \citet{Fontani2011}. Similarly we have about the same values for D$_\textrm{R}$ as Fontani (p-values $>$ 0.5). A major factor in the differences in the samples is the distance at which they are observed, which influences the filling factor. Thus higher resolution observations are needed to provide more accurate values specially in the high-mass cases which are more distant.

We tested if D$_R$ correlates with several parameters: the N$_2$H$^+$ (1--0) integrated intensity, excitation temperature, optical depth, linewidth, N$_2$D$^+$ (3--2) linewidth and N$_2$H$^+$ column density. There is no clear correlation between D$_\textrm{R}$ and any of these parameters. To quantify the linear correlation we calculated the Pearson-r parameter. The highest absolute value was 0.28 (for N$_2$H$^+$ column density), a value low enough that is not statistically significantly different from zero.

\subsection{Turbulence on smaller scales}

Within low-mass star forming regions the turbulence decreases at smaller size scales and higher density \citep{Fuller1993,Pineda2013}. A smaller beam and the higher critical density of the 3--2 transition versus the 1--0 transition makes it interesting to investigate if high-mass star-forming regions show a similar trend. 

\begin{figure}
\includegraphics[width=0.47\textwidth]{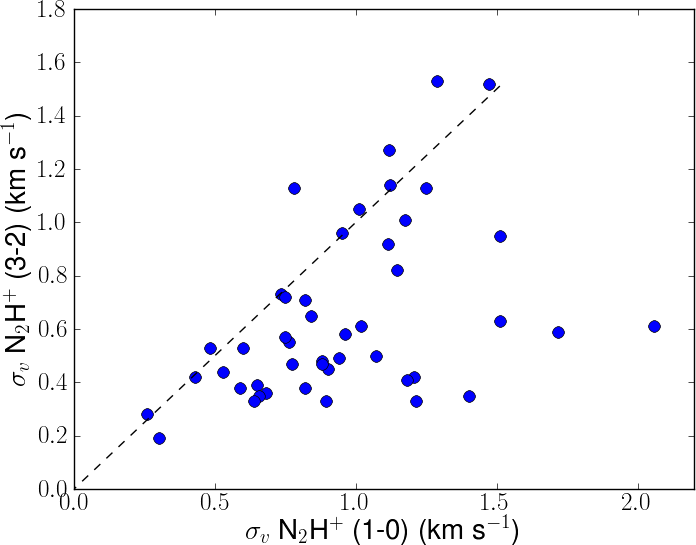}
\caption[]{Comparison of the N$_2$H$^+$ (1--0) linewidth (MOPRA) with the N$_2$H$^+$ (3--2) linewidth (APEX). The dashed line represents the identity line. }
\label{fig:sigma_vsmopra}
\end{figure}

Fig. \ref{fig:sigma_vsmopra} compares the velocity dispersions of N$_2$H$^+$ (1--0) and N$_2$H$^+$ (3--2). It shows that 16 of the 27 pointings have a smaller 3--2 linewidth. In fact, the N$_2$H$^+$ (3--2) velocity dispersion is on average $\approx 78\%$ of the N$_2$H$^+$ (1--0) dispersion. These results are consistent with \citet{Pineda2013} who show that on average the N$_2$H$^+$ (3--2) velocity dispersion is $\approx 70\%$ of the N$_2$H$^+$ (1--0) dispersion. Further analysis of Fig. \ref{fig:sigma_vsmopra} could point out three possible sub-groups: the ones that follow the 0.78 relation, the ones above the identity line (i.e. velocity dispersion is higher in the 3--2 transition), and finally a group with very narrow lines (eight objects have $< 40$\%) in the 3--2 transition.

\begin{figure}
\includegraphics[width=0.47\textwidth]{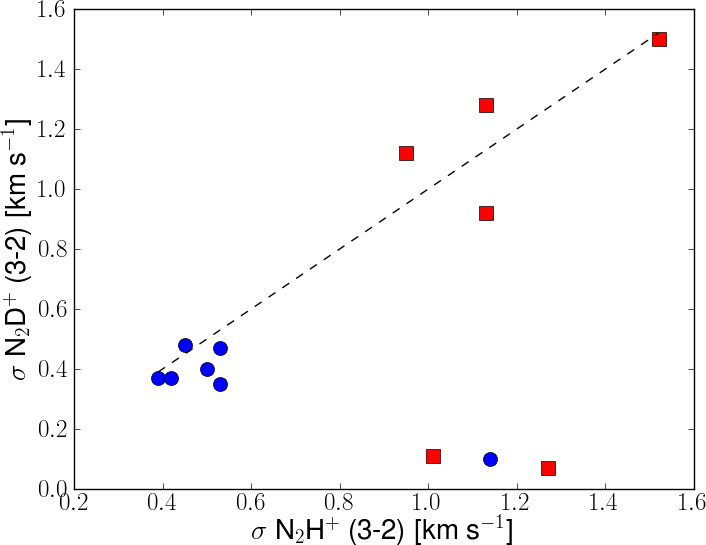}
\caption[]{Comparison of the N$_2$H$^+$ (3--2) linewidth with the N$_2$D$^+$ (3--2) linewidth. The blue dots and red squares indicate objects that exhibit one and two velocity components in the N$_2$H$^+$ (3--2) spectra, respectively. The dashed line represents the identity line.}
\label{fig:sigma_vs}
\end{figure}

Fig. \ref{fig:sigma_vs} compares the linewidth of N$_2$H$^+$ (3--2) and N$_2$D$^+$ (3--2), and it shows than N$_2$D$^+$ (3--2) exhibits an even smaller linewidth for 11 of 13 pointings. On average the N$_2$D$^+$ (3--2) velocity dispersion is $\approx 74\%$ of the N$_2$H$^+$ (3--2) dispersion, with a typical N$_2$H$^+$ (3--2) velocity dispersion of 0.63 \kms. This suggests that the N$_2$D$^+$ (3--2) traces an even smaller and more quiescent region within the beam than the N$_2$H$^+$ (3--2). In contrast, \citet{Pineda2013} found almost no difference in linewidths between N$_2$H$^+$ (3--2) and N$_2$D$^+$ (3--2), the latter only 95\% smaller.

However a more detailed analysis could suggest a different interpretation. In Fig. \ref{fig:sigma_vs} there are three objects that are clearly separated, showing very narrow lines: SDC332.26A, SDC332.26B and SDC335.283. The spectra of these objects are well fitted, but for SDC335.283 the detection is just above 3$\sigma$. The lines are so narrow that they fall below the sound speed of an average molecule ($\approx$ 0.2 \kms\ at 10 K and a mean molecular weight of 2.33). If we consider these three points as outliers, the mean ratio of linewidth between N$_2$H$^+$ (3--2) and N$_2$D$^+$ (3--2) is 96\% which is concordance with the values seen by \citet{Pineda2013}. In addition it is interesting to note in Fig. \ref{fig:sigma_vs} the clear dichotomy between objects with one or two velocity components in N$_2$H$^+$ (3--2). The two velocity component objects have much higher linewidths while following the same trend.

\section{Discussion}

\subsection{Filling factor}

The size of emission the regions of the N$_2$H$^+$ and N$_2$D$^+$ (3--2) transitions are likely to be different. The N$_2$D$^+$ region is likely to be smaller because this species should be present only in the coldest and densest parts of the N$_2$H$^+$ (3--2) region. This is consistent with high angular resolution observations.

\citet{Tan2013} observed six sources in N$_2$D$^+$ (3--2) using ALMA and obtained a mean value for the size (diameter) of these cores of 0.11 pc with a maximum value of 0.175 pc. These sizes may be a lower limits as there may be flux loss in the interferometry data leading to a bias towards small structures. For the N$_2$D$^+$ (3--2) APEX beam, and using 0.09 pc as the upper limit of the size for all the detected N$_2$D$^+$ (3--2) sources, the calculated filling factors are 0.07 to 0.35 with a mean of 0.2. We have a filling factor corrected N$_2$H$^+$ column density, applying this filling factors to the N$_2$D$^+$ column densities, the deuteration ratio D$_\textrm{R}$ increases by a factor of $\sim$ 5.0, highlighting the need of higher angular resolution to accurately constrain the deuteration in these regions.

%Even though using the N$_2$H$^+$ (3--2) observations is likely to be overestimate the real value of the column density, we obtain for reference a likely filling factor for when N$_2$H$^+$ (3--2) observations are to be used (e.g. there is lack of N$_2$H$^+$ (1--0) observations).
Interferometer observations of N$_2$H$^+$ 3--2 can also be used to estimate the filling factor. \citet{Chen2010} observed N$_2$H$^+$ (3--2) in three regions using Submillimeter Array (SMA), and determined that the N$_2$H$^+$ (3--2) was primarily in structures with size $\ga$ 0.2 pc. At the distance of our sources the APEX beam for N$_2$H$^+$ (3--2) corresponds to linear sizes of 0.24--0.55 pc. Adopting a diameter for the N$_2$H$^+$ (3--2) region of 0.2 pc we estimate beam filling factors for the detected N$_2$H$^+$ (3--2) of between 0.13 and 0.67 with a mean of 0.38. Compared with the filling factor of unity adopted above this lower filling factor would increase the deuteration ratio by a factor $\sim$ 2, highlighting again the need of high angular resolution.
%Thus if initially a filling factor of unity was used (in addition to the N$_2$D$^+$ (3--2) observation), the deuteration ratio D$_R$ increases by a factor of 1.9, highlighting again the need of high angular resolution.

\subsection{70 $\upmu$m content}

We have used the HiGal 70 $\upmu$m observations to investigate the presence of young stellar objects (YSOs) in our sample (Peretto et al., in preparation). We determine the number of compact 70 $\upmu$m sources within the APEX beam by visual inspection. In the appendix (available online) we show the 70 $\upmu$m emission for the targets with N$_2$D$^+$ (3--2) observations. 16 out of the 51 pointings have a 70 $\upmu$m source within the APEX beam. If we reduce the sample to those with N$_2$D$^+$ (3--2) detections, six pointings have a 70 $\upmu$m source within the APEX beam.

The heating of material by a protostar should rapidly reduce any enhancement in the deuterium species as the CO unfreezes in the inner parts \citep{Emprechtinger2009}. \citet{Friesen2010} sees an anti-correlation between D$_\textrm{R}$ and the distance to heating sources such as embedded protostars. So the presence of a 70 $\upmu$m source should inhibit the presence of N$_2$D$^+$. But in our sample almost half (6 out of 13) of the objects with N$_2$D$^+$ (3--2) detections have a 70 $\upmu$m source within the beam. There are two plausible scenarios for this incongruence. First there may be two distinct, different cores within beam; one with a heating source and little N$_2$D$^+$, and one without a heating source but with N$_2$D$^+$ emission. This scenario is consistent with the observations of two distinct components in the N$_2$H$^+$ (3--2) emission, and the simulations showing that substructure \citep{Smith2012}. The other possibility is that we are seeing the phase after the start of the central heating but before the heating fully suppresses the N$_2$D$^+$. So there may be a central region with no N$_2$D$^+$, and an outer zone with pockets of N$_2$D$^+$ being shielded by the central material and so still cold enough to maintain the deuterium enhancement. Distinguishing between these two possibilities require higher angular resolution observations.

\begin{figure}
\includegraphics[width=0.47\textwidth]{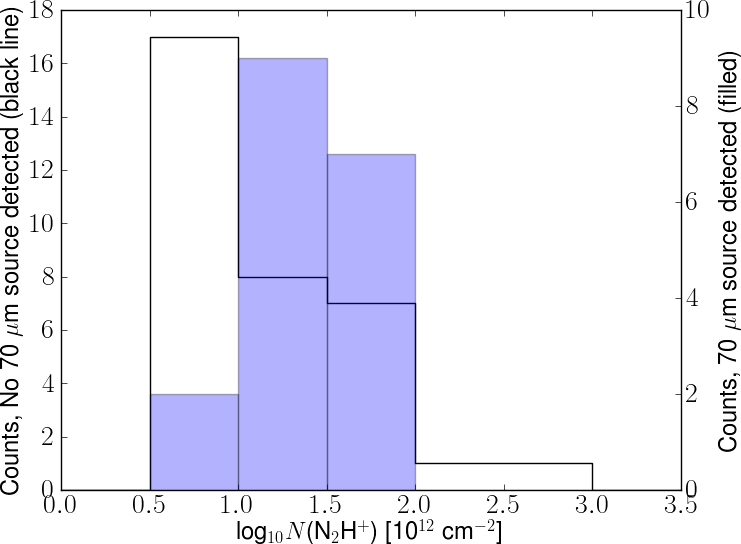}
\caption[]{Histogram of N$_2$H$^+$ column density for two subsamples: the filled histogram shows sources with a compact 70 $\upmu$m source within the APEX beam, and the black line is the histogram for sources without a 70 $\upmu$m source. }
\label{fig:histo_70um}
\end{figure}

The presence of a 70$\upmu$m source may affect the properties of the cloud, thus the properties observed in N$_2$H$^+$ (1--0). To evaluate this possibility we separated into two samples; detections and non-detections of a 70$\upmu$m source within the beam, and performed statistical tests on the obtained line properties. Kolmogorov--Smirnov tests were evaluated for N$_2$H$^+$ (1--0) velocity dispersion ($\sigma$), integrated intensity, column density, optical depth and excitation temperature (\tex). For $\sigma$ and \tex\ we cannot reject the null hypothesis at a 0.05 level. This means that the presence of a 70$\upmu$m source is not affecting the linewidth or excitation temperature. The most significant difference is in the N$_2$H$^+$ column density (Fig. \ref{fig:histo_70um}). There are few sources with column densities less than 1.0 x 10$^{12}$ cm$^{-2}$ which contain 70 $\upmu$m sources. There are however regions without 70 $\upmu$m sources with high N$_2$H$^+$ column densities similar to regions containing 70 $\upmu$m sources. These 13 sources with high column densities and no YSOs are regions of potentially high levels of deuteration, which should be investigated.
%\# observed, values obtained. 'High' n2h+ column density greater 30x1e12, and no yso. 9 observed of the 13
% 5 had n2d+ detection: fil_321B, fil351-1, fil351-2, fil351-4, L332-1
% 4 had not: L332-13, L332-15, L332-6, SDC336-2

\subsection{time-scales}

\citet{Kong2015} used a gas-phase chemical model to investigate the evolution of the deuteration ratio with time, and the impact of various parameters on the time-scale. Their results can be compared with observations to provide information about the age of the cold dense gas. Their main fiducial model used typical values of low- and high-mass pre-stellar cores. They found that D$_R$ increased with density and CO depletion factor, and decreased with temperature. In general, the model suggests that in order to reproduce the observed D$_\textrm{R}$ values time-scales of several times the free-fall time-scale are required, meaning that the observed cores are dynamically old.

We compared the \citet{Kong2015} main fiducial model with our observations, in order to evaluate the degree of evolution of our sample. We had to convert the model's N$_2$H$^+$ and N$_2$D$^+$ densities to column densities. For this we assumed a simple Gaussian density distribution with a peak value equal to the model density, and a width of 0.5 pc. 

\begin{figure}
\includegraphics[width=0.47\textwidth]{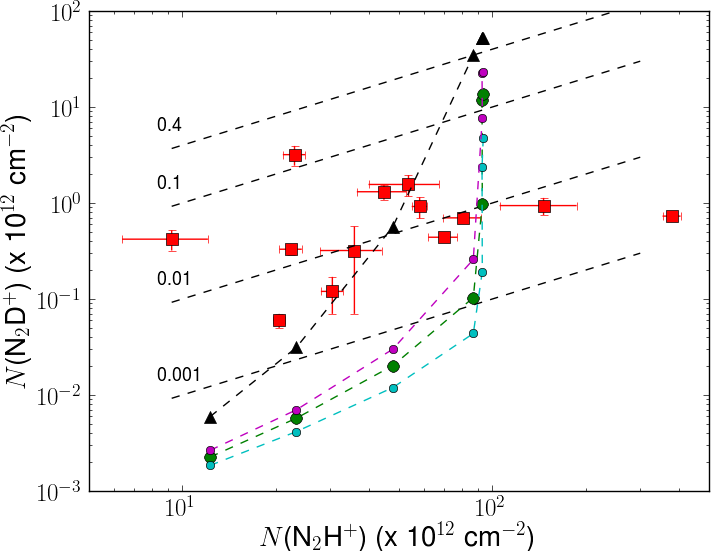}
\caption[]{N$_2$H$^+$ column density versus N$_2$D$^+$ column density. Dashed lines are constant D$_\textrm{R}$ ratio of 0.01, 0.04, and 0.25. Red squares represent our observed values. Green dots represent \citet{Kong2015} fiducial model (n = 10$^5$ cm$^{-3}$, f$_\textrm{D}$ = 10) converted values at different times, from 5 $\times$ 10$^4$ yr to 3.2 $\times$ 10$^6$ yr (each point is twice the age of the previous point). The lower and upper dot curves represent models with density of 10$^4$ cm$^{-3}$ and 10$^6$ cm$^{-3}$ respectively (f$_\textrm{D}$ = 10).  Black triangles are a model with a higher f$_D$ = 100 (n = 10$^5$ cm$^{-3}$).}
\label{fig:times}
\end{figure}

Fig. \ref{fig:times} shows a comparison of the \citet{Kong2015} models and our observations. The figure shows the column densities at different times, from 5 $\times$ 10$^4$ yr to 3.2 $\times$ 10$^6$ yr. The obtained values are only for one model, the fiducial model of their paper, with main parameter values: number density of H (n$_H$) = 10$^5$ cm$^{-3}$, depletion factor (f$_\textrm{D}$) = 10, cosmic ray ionization rate ($\zeta$) = 2.5 $\times$ 10$^{-17}$s$^{-1}$, temperature = 15 K and dust-to-gas ratio of 0.01. The values of the main model should provide a good first approximation to the objects. For example, \cite{Peretto2010} shows that the temperature of IRDCs is around 15 K. This model does a reasonable job in matching the observed values. A full exploration of the model parameters (plus size of core) should provide the whole range of observed values. The time corresponding to lowest observed D$_\textrm{R}$ is $\sim$ 5.5 $\times$ 10$^5$ years for the main model, which is about four times the free-fall time-scale for this model (1.39 $\times$ 10$^5$ yr). So this approximation to the time-scales of our sample gives dynamically old cores for even the lowest observed D$_\textrm{R}$. In addition to the main model, we consider three additional models: one with higher density, one with lower density, and one with increased depletion. As expected, given the models of \citet{Kong2015}, varying the densities has a low impact on D$_\textrm{R}$ mainly a small scaling effect. On the other hand the impact of a depletion factor 10 times higher is significant, D$_\textrm{R}$ grows much quicker. For the lowest D$_\textrm{R}$ it would only take 1.3 $\times$ 10$^5$ yr to reach the observed value which is about the free-fall time-scale suggesting the cores have a short lifetime. On the other hand, the effect of the filling factor should increase the time-scales. More detailed or complex models are necessary to further understand the time evolution of the different cores. 

\section{Summary}

In this paper we have presented the results of N$_2$H$^+$ (1--0), N$_2$H$^+$ (3--2) and N$_2$D$^+$ (3--2) observations towards peaks in IRDCs. The main results of the paper are summarized as follows.

\begin{enumerate}[i]

\item{We have observed 54 cores towards IRDCs in N$_2$H$^+$ (3--2) and 29 of them also in N$_2$D$^+$ (3--2) with APEX, most of these are also observed in N$_2$H$^+$ (1--0) with MOPRA. Our sample is the largest samples of N$_2$H$^+$ and deuterium observations towards IRDCs to date.}

\item{ 
The N$_2$H$^+$ (3--2) spectra exhibit two velocity components in 25\% of the sources. No source has two velocity components in N$_2$D$^+$ (3--2).
}

\item{N$_2$D$^+$ (3--2) was detected towards 13 objects. Its detection does not correlate with the presence of two velocity components or any other N$_2$H$^+$ (3--2) parameter except the excitation temperature. Sources detected in N$_2$D$^+$ (3--2) show a higher \tex\ suggesting a higher density.
}

\item{Modelling the N$_2$H$^+$ lines we find a filling factor of 0.4 for the N$_2$H$^+$ (1--0) transition, this implies a filling factor of unity for N$_2$H$^+$ (3--2) and N$_2$D$^+$ (3--2). We calculated the deuteration ratio, D$_\textrm{R}$, and obtained values between 0.003 and 0.14. Using alternative estimates of the N$_2$D$^+$ filling factor increases these values by a factor of at least $\sim$5.0. Compared with previous works, we observed in general higher N$_2$H$^+$ column densities, especially compared with low-mass regions, however we find similar values of D$_\textrm{R}$ compared with observations of high-mass regions by \citet{Fontani2011}.
}

\item{The measured velocity dispersions of N$_2$H$^+$ (3--2) are in general narrower ($\approx$ 80\%) than N$_2$H$^+$ (1--0) with a typical N$_2$H$^+$ (3--2) velocity dispersion 0.63 \kms. The situation comparing N$_2$D$^+$ (3--2) and N$_2$H$^+$ (3--2) is somewhat more complex as there are three outliers with very narrow N$_2$D$^+$ (3--2) velocity dispersion. For the remainder of the sources the velocity dispersions are very similar in the two species and tend to be grouped around either 0.5 \kms\ or 1.2 \kms.
}

\item{The presence of a 70$\upmu$m source is not correlated with a detection of N$_2$D$^+$ (3--2), which is surprising. In addition, the presence of a 70$\upmu$m source does not correlate with any change in linewidth and excitation temperature. However, sources with higher N$_2$H$^+$ column density are more likely to be associated with the presence of a 70$\upmu$m source.
}

\item{Comparison with the recent models of deuteration suggest evolutionary time-scales of these regions of several free-fall times or less. However the time-scales are highly dependant on the depletion factor of CO.
}

\end{enumerate}

Our observations have shown that N$_2$H$^+$ is widespread and a good tracer of the kinematics. N$_2$H$^+$ and N$_2$D$^+$ (3--2) motions are in general still supersonic even in the smallest regions traced in these observations. However, three intriguing regions appear to show subsonic turbulence in N$_2$D$^+$ in marked contrast to the N$_2$H$^+$. These regions deserve further investigation, as does the chemistry of the regions with a 70$\upmu$m source. In future more sophisticated models and higher resolution observations will help to better constrain the time-scales and evolution of these dense regions.

\section*{Acknowledgements}

M.L. acknowledges the support of CONICYT Becas-Chile 72110732. G.G. acknowledges support from CONICYT project PFB-06.

\bibliographystyle{mn2e}
\bibliography{paper}

\section*{SUPPORTING INFORMATION}

Additional Supporting Information may be found in the online version of this article:

\medskip
\textbf{Appendix\_online.pdf} \hfill \break
(\url{http://www.mnras.oxfordjournals.org/lookup/suppl/doi:10.1093/mnras/stv2354/-/DC1}).

\bigskip
This paper has been typeset from a TEX/LATEX file prepared by the author.

%\begin{thebibliography}{}
%\bibitem[\protect\citeauthoryear{{Li} \& {Nakamura}}{{Li} \&  {Nakamura}}{2006}]{Li2006}
%{Li} Z.-Y.,  {Nakamura} F.,  2006, apj, 640, L187
%\end{thebibliography}

%\clearpage

%Source & \tex\ & V$_{LSR}$ & $\sigma$ & $\tau$ & IntI & V$_{LSR}$ & $\sigma$ & $\tau$ & IntI \\ \hline
%  & K & km s$^{-1}$ & km s$^{-1}$ &   & K km s$^{-1}$ & km s$^{-1}$ & km s$^{-1}$ &   & K km s$^{-1}$ \\ \hline

%%%TABLE N2H+ 1-0

\begin{table*}
\caption{N$_2$H$^+$ (1--0) parameters. The columns show in order: the source name, kinematic distance, N$_2$H$^+$ (1--0) excitation temperature, local standard of rest velocity, linewidth, optical depth, integrated intensity, rms, and the derived column density (including the filling factor).}
\begin{tabular}{lllllllll}

\hline \hline
 & & & & N$_2$H$^+$ (1--0) & & & \\ \cline{3-8}
%Source & \tex\ & VLSR & $\sigma$ & Tau & $\int I dv$ & rms & $N$(N$_2$H$^+$) \\
% & (K) & (km s$^{-1}$) & (km s$^{-1})$ &   & (km s$^{-1}$) & (mK) & (x10$^{12}$cm$^{-2}$) \\ \hline
Source & Dist & \tex\ & V$_{\text{LSR}}$ & $\sigma$ & $\tau_0$ & $\int I dv$ & rms & $N$(N$_2$H$^+$)  \\
 & (kpc) & (K) & (km s$^{-1}$) & (km s$^{-1})$ &   & (km s$^{-1}$) & (mK) & (x10$^{12}$cm$^{-2}$) \\ \hline
SDC299.2 & 3.1 & 5.00$\pm$0.00 & -38.96$\pm$0.12 & 0.68$\pm$0.13 & 0.36$\pm$0.06 & 1.11$\pm$0.12 & 91 & 4.02$\pm$1.02 \\ 
SDC320.271 & 2.28 & 5.00$\pm$0.00 & -32.23$\pm$0.06 & 0.88$\pm$0.07 & 0.63$\pm$0.05 & 2.40$\pm$0.12 & 90 & 9.10$\pm$1.02 \\ 
SDC320.252 & 2.29 & 6.78$\pm$0.25 & -32.50$\pm$0.01 & 0.78$\pm$0.02 & 3.03$\pm$0.40 & 14.81$\pm$1.65 & 92 & 80.61$\pm$11.21 \\ 
SDC321.708 & 2.33 & 5.00$\pm$0.00 & -32.57$\pm$0.06 & 1.07$\pm$0.06 & 1.31$\pm$0.08 & 5.57$\pm$0.22 & 151 & 23.01$\pm$1.91 \\ 
SDC321.758 & 2.26 & 5.11$\pm$0.16 & -31.78$\pm$0.02 & 0.60$\pm$0.03 & 4.29$\pm$0.74 & 8.53$\pm$1.07 & 83 & 44.71$\pm$8.15 \\ 
SDC321.936 & 2.27 & 5.00$\pm$0.00 & -31.94$\pm$0.02 & 0.84$\pm$0.03 & 4.33$\pm$0.21 & 10.97$\pm$0.20 & 119 & 59.70$\pm$3.60 \\ 
SDC329.028 & 2.92 & 3.97$\pm$0.05 & -44.39$\pm$0.05 & 1.51$\pm$0.04 & 4.98$\pm$0.52 & 10.59$\pm$0.81 & 88 & 69.66$\pm$7.55 \\ 
SDC329.068 & 2.81 & 4.39$\pm$0.19 & -43.26$\pm$0.07 & 1.72$\pm$0.06 & 2.35$\pm$0.46 & 9.96$\pm$1.89 & 104 & 48.22$\pm$9.94 \\ 
SDC329.186 & 3.19 & 5.00$\pm$0.00 & -50.80$\pm$0.05 & 1.29$\pm$0.04 & 1.32$\pm$0.05 & 6.69$\pm$0.29 & 87 & 27.88$\pm$1.36 \\ 
SDC329.313 & 4.35 & 26.87$\pm$0.80 & -73.59$\pm$0.03 & 0.90$\pm$0.04 & 0.10$\pm$0.00 & 4.87$\pm$0.12 & 85 & 58.28$\pm$3.32 \\ 
SDC351.438 & 3.1 & 9.43$\pm$0.20 & -5.15$\pm$0.02 & 1.47$\pm$0.02 & 3.53$\pm$0.22 & 49.88$\pm$2.63 & 221 & 379.47$\pm$25.31 \\ 
SDC352.027 & 3.1 & 5.22$\pm$0.22 & 2.32$\pm$0.03 & 0.65$\pm$0.04 & 4.52$\pm$1.11 & 9.93$\pm$1.69 & 153 & 53.65$\pm$13.78 \\ 
SDC351.8 & 3.1 & 5.00$\pm$0.00 & -0.81$\pm$0.03 & 0.26$\pm$0.02 & 1.52$\pm$0.14 & 1.62$\pm$0.09 & 97 & 6.49$\pm$0.78 \\ 
SDC351.562 & 3.1 & 5.00$\pm$0.00 & -2.70$\pm$0.04 & 0.43$\pm$0.04 & 20.73$\pm$5.44 & 13.37$\pm$0.55 & 324 & 146.31$\pm$40.74 \\ 
L332-10 & 3.2 & 5.00$\pm$0.00 & -48.21$\pm$0.06 & 0.74$\pm$0.06 & 0.68$\pm$0.05 & 2.17$\pm$0.24 & 73 & 8.24$\pm$0.97 \\ 
L332-11 & 3.32 & 5.00$\pm$0.00 & -49.85$\pm$0.08 & 0.75$\pm$0.09 & 0.41$\pm$0.04 & 1.37$\pm$0.22 & 66 & 5.04$\pm$0.84 \\ 
L332-12 & 3.16 & 5.00$\pm$0.00 & -47.92$\pm$0.12 & 1.12$\pm$0.12 & 0.34$\pm$0.03 & 1.69$\pm$0.24 & 75 & 6.22$\pm$0.91 \\ 
L332-13 & 3.18 & 5.00$\pm$0.00 & -48.04$\pm$0.09 & 0.82$\pm$0.10 & 0.52$\pm$0.06 & 1.89$\pm$0.29 & 88 & 7.07$\pm$1.14 \\ 
L332-14 & 3.26 & 5.00$\pm$0.00 & -49.12$\pm$0.11 & 0.89$\pm$0.12 & 0.36$\pm$0.04 & 1.44$\pm$0.26 & 78 & 5.30$\pm$0.97 \\ 
L332-15 & 3.23 & 15.30$\pm$0.74 & -48.58$\pm$0.06 & 0.94$\pm$0.07 & 0.10$\pm$0.00 & 2.61$\pm$0.25 & 76 & 18.30$\pm$1.77 \\ 
L332-1 & 3.13 & 5.00$\pm$0.00 & -46.76$\pm$0.10 & 0.48$\pm$0.11 & 1.16$\pm$0.25 & 2.35$\pm$0.67 & 221 & 9.24$\pm$2.85 \\ 
L332-3 & 3.29 & 16.02$\pm$1.70 & -50.00$\pm$0.09 & -0.68$\pm$0.11 & 0.10$\pm$0.00 & 1.99$\pm$0.41 & 125 & 14.59$\pm$3.05 \\ 
L332-4 & 3.33 & 5.00$\pm$0.00 & -50.43$\pm$0.25 & 1.21$\pm$0.25 & 0.21$\pm$0.04 & 1.14$\pm$0.31 & 96 & 4.11$\pm$1.12 \\ 
L332-5 & 3.19 & 5.00$\pm$0.00 & -48.93$\pm$0.15 & 0.96$\pm$0.16 & 0.39$\pm$0.06 & 1.67$\pm$0.36 & 109 & 6.19$\pm$1.36 \\ 
L332-6 & 3.29 & 5.00$\pm$0.00 & -49.89$\pm$0.10 & 0.66$\pm$0.12 & 0.47$\pm$0.07 & 1.36$\pm$0.31 & 94 & 5.05$\pm$1.19 \\ 
L332-7 & 3.19 & 4.41$\pm$0.30 & -48.75$\pm$0.05 & 1.01$\pm$0.06 & 2.28$\pm$0.68 & 6.02$\pm$1.81 & 111 & 27.83$\pm$8.66 \\ 
L332-8 & 3.16 & 5.07$\pm$0.32 & -47.68$\pm$0.03 & 1.12$\pm$0.04 & 1.89$\pm$0.40 & 8.11$\pm$1.83 & 91 & 35.90$\pm$8.09 \\ 
L332-9 & 3.16 & 5.00$\pm$0.00 & -47.62$\pm$0.04 & 1.12$\pm$0.03 & 1.11$\pm$0.04 & 5.07$\pm$0.21 & 62 & 20.53$\pm$0.91 \\ 
SDC335.283A & 3.13 & 17.47$\pm$0.75 & -44.87$\pm$0.07 & 1.18$\pm$0.07 & 0.10$\pm$0.00 & 3.84$\pm$0.31 & 95 & 30.48$\pm$2.49 \\ 
SDC335.283B & 3.09 & 11.13$\pm$1.12 & -44.75$\pm$0.14 & 0.95$\pm$0.16 & 0.10$\pm$0.00 & 1.73$\pm$0.38 & 114 & 9.18$\pm$2.02 \\ 
SDC335.077A & 2.87 & 14.75$\pm$0.62 & -39.52$\pm$0.08 & 1.25$\pm$0.08 & 0.10$\pm$0.00 & 3.30$\pm$0.27 & 83 & 22.43$\pm$1.85 \\ 
SDC335.077B & 3.1 & 9.14$\pm$0.65 & -38.55$\pm$0.15 & 1.19$\pm$0.14 & 0.10$\pm$0.00 & 1.64$\pm$0.26 & 82 & 7.41$\pm$1.20 \\ 
SDC335.579A & 3.19 & 4.32$\pm$0.16 & -46.38$\pm$0.04 & 1.15$\pm$0.04 & 2.52$\pm$0.44 & 6.88$\pm$1.19 & 78 & 33.06$\pm$6.09 \\ 
SDC335.579B & 3.28 & 5.00$\pm$0.00 & -47.10$\pm$0.10 & 0.88$\pm$0.11 & 0.39$\pm$0.04 & 1.52$\pm$0.24 & 73 & 5.61$\pm$0.91 \\ 
SDC335.579Y1 & 3.26 & 5.02$\pm$0.19 & -46.44$\pm$0.03 & 1.21$\pm$0.03 & 2.32$\pm$0.32 & 10.00$\pm$1.39 & 84 & 46.54$\pm$6.84 \\ 
SDC335.579Y2 & 3.21 & 4.80$\pm$0.19 & -46.44$\pm$0.03 & 1.18$\pm$0.03 & 2.32$\pm$0.35 & 8.83$\pm$1.34 & 81 & 40.95$\pm$6.51 \\ 
SDC335.44 & 3.11 & 8.23$\pm$0.62 & -45.18$\pm$0.32 & 2.06$\pm$0.25 & 0.10$\pm$0.00 & 2.41$\pm$0.40 & 102 & 10.04$\pm$1.70 \\ 
SDC335.253 & 3.1 & 5.01$\pm$0.00 & -45.22$\pm$0.17 & 0.78$\pm$0.19 & 0.40$\pm$0.08 & 1.40$\pm$0.44 & 130 & 5.17$\pm$1.66 \\ 
SDC335.229A & 3.11 & 12.33$\pm$0.94 & -40.12$\pm$0.17 & 1.40$\pm$0.16 & 0.10$\pm$0.00 & 2.94$\pm$0.44 & 137 & 17.01$\pm$2.60 \\ 
SDC335.229B & 2.88 & - & - & - & - & - & - & - \\ 
SDC335.059 & 2.84 & 15.28$\pm$1.85 & -39.30$\pm$0.10 & -0.61$\pm$0.11 & 0.10$\pm$0.00 & 1.68$\pm$0.39 & 122 & 11.78$\pm$2.82 \\ 
SDC356.84 & 3.44 & 5.00$\pm$0.00 & -7.94$\pm$0.05 & 0.82$\pm$0.06 & 1.10$\pm$0.08 & 3.71$\pm$0.15 & 110 & 14.81$\pm$1.53 \\ 
SDC13.174A & 3.6 & 5.00$\pm$0.00 & 37.12$\pm$0.06 & 0.77$\pm$0.06 & 0.94$\pm$0.07 & 3.04$\pm$0.31 & 93 & 11.93$\pm$1.30 \\ 
SDC13.174B & 3.1 & 5.00$\pm$0.00 & 37.36$\pm$0.11 & 0.54$\pm$0.12 & 0.38$\pm$0.08 & 0.92$\pm$0.28 & 90 & 3.38$\pm$1.05 \\ 
SDC13.194 & 3.57 & 5.00$\pm$0.00 & 36.41$\pm$0.11 & 0.76$\pm$0.12 & 0.43$\pm$0.06 & 1.47$\pm$0.31 & 92 & 5.45$\pm$1.17 \\ 
SDC16.915 & 3.41 & 5.00$\pm$0.00 & 41.30$\pm$0.04 & 0.64$\pm$0.04 & 2.42$\pm$0.18 & 5.69$\pm$0.19 & 140 & 25.42$\pm$2.47 \\ 
SDC2.141 & 2.68 & 3.96$\pm$0.06 & 4.66$\pm$0.04 & 0.53$\pm$0.04 & 10.68$\pm$2.55 & 6.34$\pm$0.60 & 121 & 52.34$\pm$13.13 \\ 
SDC2.116 & 2.36 & 5.00$\pm$0.00 & 3.88$\pm$0.05 & 0.75$\pm$0.06 & 1.45$\pm$0.11 & 4.29$\pm$0.18 & 134 & 17.85$\pm$1.97 \\ 
SDC301.876 & 3.1 & 3.43$\pm$0.05 & -43.15$\pm$0.03 & 0.30$\pm$0.03 & 17.17$\pm$5.18 & 2.81$\pm$0.32 & 92 & 33.03$\pm$10.51 \\ 
SDC327.894 & 2.93 & 3.72$\pm$0.08 & -44.48$\pm$0.04 & 0.59$\pm$0.05 & 6.33$\pm$1.76 & 4.20$\pm$0.67 & 93 & 29.36$\pm$8.56 \\ 
SDC327.964 & 2.97 & 5.00$\pm$0.00 & -44.85$\pm$0.03 & 0.42$\pm$0.04 & 1.05$\pm$0.09 & 1.89$\pm$0.08 & 73 & 7.24$\pm$0.93 \\ 
SDC34.558 & 3.33 & 18.15$\pm$1.37 & 53.19$\pm$0.17 & 1.51$\pm$0.14 & 0.10$\pm$0.00 & 5.17$\pm$0.38 & 267 & 42.49$\pm$5.64 \\ 
SDC37.08 & 5.13 & 5.00$\pm$0.00 & 81.70$\pm$0.06 & 1.02$\pm$0.06 & 1.57$\pm$0.10 & 6.19$\pm$0.24 & 168 & 26.29$\pm$2.28 \\   \hline
\end{tabular}
\\
\begin{flushleft}
\end{flushleft}
\label{tab:pars_n2h10}
\end{table*}

%SDC335.229B & 2.88 & - & - & - & - & - & - & - \\ 

%%%TABLE N2H+ 3-2 part 1

\begin{table*}
\caption{N$_2$H$^+$ (3--2) parameters. The columns show in order: the source name, the N$_2$H$^+$ (3--2) excitation temperature, local standard of rest velocity, linewidth, optical depth, integrated intensity, rms, and the derived column density.}
\begin{tabular}{llllllll}

\hline \hline
 & & & & N$_2$H$^+$ (3--2) & & & \\ \cline{2-7}
Source & \tex\ & V$_{\text{LSR}}$ & $\sigma$ & $\tau_0$ & $\int I dv$ & rms & $N$(N$_2$H$^+$) \\
 & (K) & (km s$^{-1}$) & (km s$^{-1})$ &   & (km s$^{-1}$) & (mK) & (x10$^{12}$cm$^{-2}$) \\ \hline

SDC299.2 & 6.50$\pm$0.00 & -39.69$\pm$0.03 & 0.36$\pm$0.03 & 0.24$\pm$0.02 & 0.39$\pm$0.03 & 68 & 1.48$\pm$0.17 \\ 
SDC320.271 & 3.66$\pm$0.04 & -32.44$\pm$0.05 & 0.48$\pm$0.04 & 9.75$\pm$2.30 & 0.77$\pm$0.06 & 52 & 219.32$\pm$74.64 \\ 
SDC320.252-a & 9.22$\pm$0.00 & -32.36$\pm$0.01 & 1.13$\pm$0.01 & 1.83$\pm$0.03 & 12.55$\pm$0.28 & 203 & 29.97$\pm$0.56 \\ 
SDC320.252-b & 9.22$\pm$0.00 & -35.72$\pm$0.02 & 0.52$\pm$0.02 & 0.42$\pm$0.01 & 1.94$\pm$0.28 & 203 & 3.17$\pm$0.14 \\ 
SDC321.708 & 4.47$\pm$0.03 & -33.50$\pm$0.02 & 0.50$\pm$0.02 & 6.27$\pm$0.95 & 1.61$\pm$0.08 & 76 & 91.11$\pm$17.67 \\ 
SDC321.758 & 5.57$\pm$0.04 & -32.29$\pm$0.02 & 0.53$\pm$0.02 & 6.93$\pm$0.78 & 3.60$\pm$0.12 & 117 & 74.20$\pm$11.32 \\ 
SDC321.936 & 8.48$\pm$0.02 & -32.52$\pm$0.01 & 0.65$\pm$0.01 & 14.20$\pm$0.38 & 14.91$\pm$0.10 & 325 & 135.77$\pm$4.94 \\ 
SDC329.028-a & 11.12$\pm$0.00 & -42.00$\pm$0.01 & 0.95$\pm$0.01 & 2.61$\pm$0.05 & 18.39$\pm$1.33 & 155 & 36.46$\pm$0.80 \\ 
SDC329.028-b & 9.37$\pm$0.22 & -46.41$\pm$0.02 & 1.44$\pm$0.04 & 1.92$\pm$0.23 & 16.81$\pm$1.33 & 155 & 40.02$\pm$8.38 \\ 
SDC329.068-a & 4.61$\pm$0.06 & -40.59$\pm$0.05 & 0.59$\pm$0.03 & 17.48$\pm$2.73 & 2.94$\pm$0.53 & 95 & 282.10$\pm$76.15 \\ 
SDC329.068-b & 4.35$\pm$0.06 & -43.66$\pm$0.09 & 0.95$\pm$0.12 & 7.33$\pm$3.01 & 2.64$\pm$0.53 & 95 & 59.44$\pm$21.22 \\ 
SDC329.186-a & 6.46$\pm$0.20 & -50.73$\pm$0.08 & 1.53$\pm$0.10 & 4.72$\pm$1.19 & 11.62$\pm$2.31 & 144 & 123.97$\pm$54.01 \\ 
SDC329.186-b & 7.51$\pm$0.14 & -48.68$\pm$0.04 & 1.21$\pm$0.04 & 6.23$\pm$0.69 & 14.65$\pm$2.31 & 144 & 116.30$\pm$24.52 \\ 
SDC329.313 & 5.38$\pm$0.03 & -73.87$\pm$0.02 & 0.45$\pm$0.01 & 17.53$\pm$1.24 & 3.96$\pm$0.10 & 133 & 167.08$\pm$17.91 \\ 
SDC351.438-a & 14.99$\pm$0.00 & -6.07$\pm$0.04 & 1.52$\pm$0.03 & 1.95$\pm$0.03 & 39.77$\pm$1.56 & 387 & 50.10$\pm$1.25 \\ 
SDC351.438-b & 14.99$\pm$0.00 & -1.64$\pm$0.03 & 1.08$\pm$0.01 & 2.26$\pm$0.07 & 31.13$\pm$1.56 & 387 & 41.26$\pm$1.33 \\ 
SDC352.027 & 4.96$\pm$0.04 & 1.97$\pm$0.02 & 0.39$\pm$0.01 & 10.65$\pm$1.21 & 2.29$\pm$0.09 & 96 & 99.68$\pm$16.85 \\ 
SDC351.8 & 6.50$\pm$0.00 & -1.18$\pm$0.03 & 0.28$\pm$0.03 & 0.25$\pm$0.02 & 0.31$\pm$0.02 & 51 & 1.20$\pm$0.16 \\ 
SDC351.562 & 5.56$\pm$0.03 & -3.13$\pm$0.01 & 0.42$\pm$0.01 & 13.53$\pm$0.92 & 3.74$\pm$0.09 & 122 & 115.08$\pm$11.74 \\ 
L332-10 & 7.72$\pm$0.33 & -48.32$\pm$0.08 & 0.73$\pm$0.08 & 0.10$\pm$0.00 & 0.49$\pm$0.04 & 71 & 1.08$\pm$0.53 \\ 
L332-11 & 6.50$\pm$0.00 & -50.94$\pm$0.05 & 0.72$\pm$0.05 & 0.25$\pm$0.02 & 0.76$\pm$0.04 & 70 & 3.07$\pm$0.33 \\ 
L332-12 & 9.38$\pm$0.26 & -47.57$\pm$0.06 & 0.92$\pm$0.06 & 0.10$\pm$0.00 & 0.92$\pm$0.04 & 71 & 1.29$\pm$0.33 \\ 
L332-13 & 4.76$\pm$0.07 & -48.06$\pm$0.03 & 0.38$\pm$0.03 & 5.30$\pm$1.41 & 1.47$\pm$0.14 & 122 & 51.91$\pm$18.83 \\ 
L332-14 & 4.12$\pm$0.05 & -49.55$\pm$0.03 & 0.33$\pm$0.03 & 7.66$\pm$2.02 & 0.88$\pm$0.08 & 71 & 87.64$\pm$31.44 \\ 
L332-15 & 5.19$\pm$0.04 & -48.86$\pm$0.02 & 0.49$\pm$0.02 & 11.29$\pm$1.15 & 3.24$\pm$0.10 & 113 & 123.51$\pm$19.31 \\ 
L332-1 & 6.32$\pm$0.03 & -46.81$\pm$0.02 & 0.53$\pm$0.01 & 17.31$\pm$0.88 & 6.99$\pm$0.12 & 167 & 160.72$\pm$12.40 \\ 
L332-3 & 5.84$\pm$0.04 & -50.19$\pm$0.01 & 0.38$\pm$0.01 & 14.48$\pm$1.04 & 4.05$\pm$0.10 & 149 & 104.83$\pm$12.12 \\ 
L332-4 & 6.50$\pm$0.00 & -50.93$\pm$0.03 & 0.33$\pm$0.03 & 0.58$\pm$0.05 & 0.77$\pm$0.05 & 121 & 3.27$\pm$0.41 \\ 
L332-5 & 6.50$\pm$0.00 & -47.98$\pm$0.05 & 0.58$\pm$0.05 & 0.24$\pm$0.02 & 0.60$\pm$0.04 & 68 & 2.38$\pm$0.28 \\ 
L332-6 & 4.88$\pm$0.05 & -50.08$\pm$0.02 & 0.35$\pm$0.02 & 9.13$\pm$1.50 & 1.88$\pm$0.10 & 122 & 78.83$\pm$18.55 \\ 
L332-7-a & 10.89$\pm$0.00 & -48.03$\pm$0.01 & 1.05$\pm$0.01 & 2.35$\pm$0.03 & 18.49$\pm$5.61 & 236 & 36.12$\pm$0.58 \\ 
L332-7-b & 6.21$\pm$1.22 & -51.42$\pm$0.05 & 0.75$\pm$0.09 & 0.96$\pm$0.96 & 2.26$\pm$5.61 & 236 & 9.74$\pm$36.64 \\ 
L332-8-a & 8.83$\pm$0.00 & -46.83$\pm$0.09 & 1.27$\pm$0.04 & 2.84$\pm$0.12 & 16.16$\pm$2.34 & 111 & 52.58$\pm$2.77 \\ 
L332-8-b & 6.52$\pm$0.15 & -50.54$\pm$0.13 & 0.92$\pm$0.06 & 3.02$\pm$0.67 & 6.37$\pm$2.34 & 111 & 47.31$\pm$16.35 \\ 
L332-9 & 22.91$\pm$0.32 & -47.58$\pm$0.02 & 1.14$\pm$0.02 & 0.10$\pm$0.00 & 4.64$\pm$0.08 & 101 & 2.75$\pm$0.13 \\ 
SDC335.283A-a & 8.21$\pm$0.00 & -44.30$\pm$0.06 & 1.01$\pm$0.02 & 2.37$\pm$0.12 & 10.34$\pm$2.89 & 130 & 35.55$\pm$1.93 \\ 
SDC335.283A-b & 5.07$\pm$0.07 & -47.39$\pm$0.16 & 0.82$\pm$0.07 & 4.43$\pm$1.24 & 3.36$\pm$2.89 & 130 & 84.12$\pm$30.04 \\ 
SDC335.283B & 6.50$\pm$0.00 & -43.97$\pm$0.05 & 0.96$\pm$0.05 & 0.44$\pm$0.02 & 1.71$\pm$0.07 & 101 & 7.21$\pm$0.50 \\ 
SDC335.077A-a & 10.40$\pm$0.00 & -39.40$\pm$0.03 & 1.13$\pm$0.02 & 1.33$\pm$0.04 & 12.92$\pm$9.15 & 352 & 21.83$\pm$0.76 \\ 
SDC335.077A-b & 5.59$\pm$0.08 & -41.34$\pm$0.06 & 0.92$\pm$0.05 & 7.33$\pm$1.45 & 5.81$\pm$9.15 & 352 & 135.61$\pm$37.94 \\ 
SDC335.077B & - & - & - & - & - & 40 & 0.7$^a$ \\ 
SDC335.579A-a & 5.94$\pm$0.00 & -44.80$\pm$0.05 & 0.82$\pm$0.04 & 1.68$\pm$0.10 & 3.25$\pm$1.31 & 140 & 25.74$\pm$1.98 \\ 
SDC335.579A-b & 5.23$\pm$0.16 & -47.11$\pm$0.04 & 0.45$\pm$0.04 & 3.03$\pm$1.04 & 1.79$\pm$1.31 & 140 & 15.46$\pm$13.96 \\ 
SDC335.579B & 4.29$\pm$0.03 & -47.71$\pm$0.03 & 0.47$\pm$0.02 & 16.19$\pm$1.79 & 1.83$\pm$0.07 & 77 & 240.98$\pm$41.41 \\ 
SDC335.579Y1-a & 6.86$\pm$0.04 & -44.19$\pm$0.01 & 0.42$\pm$0.01 & 17.09$\pm$0.87 & 7.10$\pm$0.24 & 375 & 117.29$\pm$9.76 \\ 
SDC335.579Y1-b & 12.31$\pm$0.03 & -47.58$\pm$0.01 & 0.79$\pm$0.01 & 9.36$\pm$0.22 & 30.27$\pm$0.24 & 375 & 112.36$\pm$3.30 \\ 
SDC335.579Y2-a & 8.10$\pm$0.05 & -44.19$\pm$0.01 & 0.41$\pm$0.01 & 10.58$\pm$0.46 & 8.34$\pm$0.41 & 257 & 64.72$\pm$4.72 \\ 
SDC335.579Y2-b & 10.73$\pm$0.00 & -47.44$\pm$0.01 & 1.43$\pm$0.01 & 1.49$\pm$0.02 & 18.44$\pm$0.41 & 257 & 31.10$\pm$0.47 \\ 
SDC335.44 & 4.90$\pm$0.04 & -44.35$\pm$0.04 & 0.61$\pm$0.02 & 16.60$\pm$1.76 & 3.65$\pm$0.11 & 121 & 248.06$\pm$41.59 \\ 
SDC335.253 & - & - & - & - & - & 41 & 0.71$^a$ \\ 
SDC335.229A-a & 3.39$\pm$0.06 & -40.07$\pm$0.08 & 0.35$\pm$0.06 & 13.09$\pm$6.11 & 0.43$\pm$0.13 & 68 & 62.08$\pm$37.36 \\ 
SDC335.229A-b & 3.85$\pm$0.00 & -44.46$\pm$0.06 & 0.39$\pm$0.06 & 1.54$\pm$0.34 & 0.34$\pm$0.13 & 68 & 24.58$\pm$6.61 \\ 
SDC335.229B & 6.50$\pm$0.00 & -39.92$\pm$0.07 & 0.63$\pm$0.06 & 0.27$\pm$0.03 & 0.72$\pm$0.06 & 102 & 2.90$\pm$0.42 \\ 
\hline

\end{tabular}
\label{tab:pars_n2h}
\end{table*}

%%%TABLE N2H+ 3-2 part 2

\begin{table*}
\caption{N$_2$H$^+$ (3--2) parameters continued.}
\begin{tabular}{llllllll}

\hline \hline
 & & & & N$_2$H$^+$ (3--2) & (continuation) & & \\ \cline{2-7}
Source & \tex\ & V$_{\text{LSR}}$ & $\sigma$ & $\tau_0$ & $\int I dv$ & rms & $N$(N$_2$H$^+$) \\
 & (K) & (km s$^{-1}$) & (km s$^{-1})$ &   & (km s$^{-1}$) & (mK) & (x10$^{12}$cm$^{-2}$) \\ \hline

SDC335.059 & 6.50$\pm$0.00 & -39.19$\pm$0.20 & 0.66$\pm$0.20 & 0.05$\pm$0.01 & 0.16$\pm$0.04 & 38 & 0.58$\pm$0.14 \\ 
SDC356.84 & 6.50$\pm$0.00 & -8.48$\pm$0.06 & 0.71$\pm$0.06 & 0.22$\pm$0.02 & 0.67$\pm$0.04 & 57 & 2.67$\pm$0.33 \\ 
SDC13.174A & 4.47$\pm$0.05 & 36.90$\pm$0.03 & 0.47$\pm$0.03 & 6.24$\pm$1.39 & 1.53$\pm$0.11 & 80 & 85.23$\pm$25.56 \\ 
SDC13.174B & - & - & - & - & - & 99 & 1.72$^a$ \\ 
SDC13.194 & 6.50$\pm$0.00 & 36.42$\pm$0.04 & 0.55$\pm$0.04 & 0.31$\pm$0.02 & 0.73$\pm$0.04 & 63 & 2.91$\pm$0.28 \\ 
SDC16.915 & 4.76$\pm$0.05 & 40.90$\pm$0.02 & 0.33$\pm$0.01 & 16.75$\pm$2.06 & 2.04$\pm$0.10 & 109 & 142.47$\pm$29.86 \\ 
SDC2.141 & 6.50$\pm$0.00 & 4.27$\pm$0.02 & 0.44$\pm$0.02 & 0.55$\pm$0.03 & 0.96$\pm$0.04 & 155 & 4.13$\pm$0.29 \\ 
SDC2.116 & 6.50$\pm$0.00 & 3.52$\pm$0.03 & 0.57$\pm$0.03 & 0.34$\pm$0.02 & 0.83$\pm$0.04 & 70 & 3.31$\pm$0.26 \\ 
SDC301.876 & 4.27$\pm$0.06 & -43.27$\pm$0.02 & 0.19$\pm$0.02 & 9.20$\pm$2.37 & 0.70$\pm$0.06 & 75 & 55.92$\pm$20.97 \\ 
SDC327.894 & 3.95$\pm$0.04 & -44.52$\pm$0.03 & 0.38$\pm$0.02 & 11.74$\pm$2.33 & 0.97$\pm$0.07 & 70 & 171.10$\pm$48.69 \\ 
SDC327.964 & 6.50$\pm$0.00 & -45.23$\pm$0.07 & 0.50$\pm$0.07 & 0.14$\pm$0.02 & 0.32$\pm$0.03 & 66 & 1.20$\pm$0.24 \\ 
SDC34.558 & 6.50$\pm$0.00 & 52.73$\pm$0.03 & 0.63$\pm$0.03 & 0.36$\pm$0.02 & 0.95$\pm$0.04 & 73 & 3.87$\pm$0.28 \\ 
SDC37.08 & 5.42$\pm$0.03 & 81.50$\pm$0.02 & 0.61$\pm$0.02 & 15.43$\pm$1.12 & 4.87$\pm$0.10 & 138 & 197.29$\pm$21.75 \\ \hline
%\multicolumn{4}{l}{\textsuperscript{*} \footnotesize{Upper limit}}
\end{tabular}
\\
\begin{flushleft}
\textit{Note.} $^a$denotes upper limit.
\end{flushleft}
\label{tab:pars_n2h_p2}
\end{table*}

%%%TABLE N2D+

\begin{table*}
\caption{N$_2$D$^+$ (3--2) parameters. The columns show in order: the source name, the N$_2$D$^+$ (3--2) local standard of rest velocity, linewidth, optical depth, integrated intensity, rms, the derived column density, and the deuteration ratio}
\begin{tabular}{lllllrlr}
\hline \hline
 & & & N$_2$D$^+$ (3--2) & \\ \cline{2-6}
Source & V$_{\text{LSR}}$ & $\sigma$ & $\tau_0$ & $\int I dv$ & $rms$ & $N$(N$_2$D$^+$) & D$_R$ \\
 & (K) & (km s$^{-1})$ &  & (km s$^{-1})$ & (mK) & (x10$^{12}$cm$^{-2}$) &  \\ \hline

SDC320.252-a & -32.60$\pm$0.17 & 1.28$\pm$0.17 & 0.03$\pm$0.00 & 0.42$\pm$0.04 & 34 & 0.70$\pm$0.08 & 0.009 \\ 
SDC321.708 & -32.75$\pm$0.07 & 0.40$\pm$0.07 & 0.41$\pm$0.07 & 0.31$\pm$0.04 & 39 & 3.19$\pm$0.74 & 0.139 \\ 
SDC321.758 & -31.22$\pm$0.06 & 0.47$\pm$0.06 & 0.18$\pm$0.02 & 0.31$\pm$0.03 & 33 & 1.31$\pm$0.23 & 0.029 \\ 
SDC329.028-a & -43.66$\pm$0.17 & 1.12$\pm$0.17 & 0.02$\pm$0.00 & 0.35$\pm$0.04 & 32 & 0.44$\pm$0.05 & 0.006 \\ 
SDC329.068-a & - & - & - & - & 36 & 1.01$^a$ & 0.021 \\ 
SDC329.186-a & - & - & - & - & 53 & 0.28$^a$ & 0.010 \\ 
SDC329.313 & -73.42$\pm$0.12 & 0.48$\pm$0.13 & 0.11$\pm$0.03 & 0.19$\pm$0.04 & 41 & 0.92$\pm$0.23 & 0.016 \\ 
SDC351.438-a & -3.88$\pm$0.11 & 1.50$\pm$0.12 & 0.02$\pm$0.00 & 0.88$\pm$0.05 & 35 & 0.73$\pm$0.04 & 0.002 \\ 
SDC352.027 & 2.50$\pm$0.07 & 0.37$\pm$0.08 & 0.24$\pm$0.05 & 0.24$\pm$0.03 & 32 & 1.56$\pm$0.39 & 0.029 \\ 
SDC351.562 & -2.64$\pm$0.08 & 0.37$\pm$0.09 & 0.16$\pm$0.03 & 0.22$\pm$0.03 & 39 & 0.94$\pm$0.19 & 0.006 \\ 
L332-10 & - & - & - & - & 67 & 0.48$^a$ & 0.058 \\ 
L332-13 & - & - & - & - & 39 & 1.23$^a$ & 0.174 \\ 
L332-15 & - & - & - & - & 55 & 1.22$^a$ & 0.067 \\ 
L332-1 & -46.66$\pm$0.10 & 0.35$\pm$0.12 & 0.08$\pm$0.02 & 0.15$\pm$0.03 & 40 & 0.42$\pm$0.10 & 0.045 \\ 
L332-3 & - & - & - & - & 59 & 0.87$^a$ & 0.060 \\ 
L332-6 & - & - & - & - & 41 & 1.16$^a$ & 0.230 \\ 
L332-7a & - & - & - & - & 58 & 0.27$^a$ & 0.010 \\ 
L332-8a & -47.75$\pm$0.03 & 0.07$\pm$0.04 & 0.28$\pm$0.37 & 0.20$\pm$0.15 & 24 & 0.32$\pm$0.25 & 0.009 \\ 
L332-9 & -46.72$\pm$0.05 & 0.10$\pm$0.06 & 0.02$\pm$0.01 & 0.09$\pm$0.02 & 37 & 0.06$\pm$0.01 & 0.003 \\ 
SDC335.283A-a & -44.36$\pm$0.08 & 0.11$\pm$0.04 & 0.06$\pm$0.02 & 0.07$\pm$0.02 & 37 & 0.12$\pm$0.05 & 0.004 \\ 
SDC335.077A-a & -39.14$\pm$0.15 & 0.92$\pm$0.14 & 0.02$\pm$0.00 & 0.27$\pm$0.03 & 32 & 0.33$\pm$0.04 & 0.015 \\ 
SDC335.579A-a & - & - & - & - & 43 & 0.88$^a$ & 0.027 \\ 
SDC335.579Y1-a & - & - & - & - & 39 & 0.14$^a$ & 0.003 \\ 
SDC335.579Y2-a & - & - & - & - & 38 & 0.20$^a$ & 0.005 \\ 
SDC335.44 & - & - & - & - & 42 & 1.17$^a$ & 0.117 \\ 
SDC335.229A-a & - & - & - & - & 37 & 0.40$^a$ & 0.024 \\ 
SDC13.174A & - & - & - & - & 58 & 2.42$^a$ & 0.203 \\ 
SDC34.558 & - & - & - & - & 43 & 0.46$^a$ & 0.011 \\ 
SDC37.08 & - & - & - & - & 37 & 0.70$^a$ & 0.027 \\ \hline
\end{tabular}
\\
\begin{flushleft}
\textit{Note.} $^a$denotes upper limit.
\end{flushleft}
\label{tab:pars_n2d}
\end{table*}

\end{document}